\documentclass[aps,pra,twocolumn,preprintnumbers,groupedaddress,amsmath,amssymb,showkeys]{revtex4-2}
\usepackage[T1]{fontenc} 
\usepackage{mathtools}
\usepackage{bm}
\usepackage{dsfont}
\usepackage{xcolor}
\usepackage{physics}
\usepackage{hyperref}
\usepackage{tikz}
\usepackage[detect-all]{siunitx}
\usepackage[shortlabels]{enumitem}

\newcommand{\pidx}[1]{{\mbox{\tiny $(#1)$}}}

\makeatletter
\def\convertto#1#2{\strip@pt\dimexpr #2*65536/\number\dimexpr 1#1}
\makeatother

\usepackage[normalem]{ulem}

\DeclareMathOperator*{\std}{std}
\DeclareMathOperator*{\argmin}{arg\,min}

\definecolor{pltblue}{HTML}{1F77B4}

\definecolor{csky}{HTML}{375E97}
\definecolor{csunset}{HTML}{FB6542}
\definecolor{csunflower}{HTML}{FFBB00}
\definecolor{cgrass}{HTML}{3F681C}

\definecolor{cblue}{HTML}{4D85BD}
\definecolor{cleafygreen}{HTML}{7CAA2D}
\definecolor{csunflower2}{HTML}{F5E356}
\definecolor{cseeds}{HTML}{CB6318}

\definecolor{csunshine}{HTML}{F9BA32}

\definecolor{lightcoral}{HTML}{F08080}

\newcommand{\Ntrain}{N_\mathrm{train}}
\newcommand{\Ntest}{N_\mathrm{test}}

\begin{document}
	
	
	\title{Photonic kernel machine learning for ultrafast spectral analysis}
	
	
	\author{Zakari Denis}
	\affiliation{Universit\'e de Paris, CNRS, Laboratoire Mat\'eriaux et Ph\'enom\`enes Quantiques, F-75013 Paris, France}
	\author{Ivan Favero}
	\affiliation{Universit\'e de Paris, CNRS, Laboratoire Mat\'eriaux et Ph\'enom\`enes Quantiques, F-75013 Paris, France}
	\author{Cristiano Ciuti}
	\email{cristiano.ciuti@u-paris.fr}
	\affiliation{Universit\'e de Paris, CNRS, Laboratoire Mat\'eriaux et Ph\'enom\`enes Quantiques, F-75013 Paris, France}
	
	
	\date{\today}
	
	\begin{abstract}
		We introduce photonic kernel machines, a scheme for ultrafast spectral analysis of noisy radio-frequency signals from single-shot optical intensity measurements. The approach combines the versatility of machine learning and the speed of photonic hardware to reach unprecedented throughput rates. We theoretically describe some of the key underlying principles, and then numerically illustrate the reached performances on a photonic lattice-based implementation. We apply the technique both to picosecond pulsed radio-frequency signals, on energy-spectral-density estimation and a shape classification task, and to continuous signals, on a frequency tracking task. The presented optical computing scheme is resilient to noise while requiring minimal control on the photonic-lattice parameters, making it readily implementable in realistic state-of-the-art photonic platforms.
	\end{abstract}
	

\maketitle

\section{Introduction\label{sec:1}}

    As a result of its intrinsically faster timescales, photonics was very soon envisioned as a promising tool to outperform integrated electronics in terms of data processing rates~\cite{ambs2010}. In this perspective, optical setups ranging from matrix-vector multipliers~\cite{mengert66}, function convolvers \cite{weaver1966} and discrete Fourier-transforming processors~\cite{goodman1978} to non-von Neumann parallel digital processors~\cite{sawchuk1984,huang1984} were proposed. Although these ideas quickly became outdated as a consequence of the fast rise of silicon-based electronic processors, the latter have started to exhibit some of their limitations. In particular, the emergence of machine learning applications operating on ever increasing amounts of data, involving deeper and deeper neural-network architectures with an increasing degree of complexity~\cite{khan2020} has led to a situation where the progress of available digital processor technology no longer keeps pace with the demand in computing capabilities~\cite{waldrop2016}. This trend has moreover gone hand in hand with an increase in the consumption of computational and energy resources, casting doubt on its sustainability~\cite{strubell2019}.
    
    This context has stimulated very interesting proposals aiming at surrogating the realization of specific tasks that are computationally and energetically very demanding to very specialized (electro-)optical devices. This process, reminiscent of current trends in hardware acceleration technologies, such as graphical (GPUs) and tensor processing units (TPUs), has already led to commercially available optical co-processors~\cite{dong2018,ohana2020,schneider2020,schneider2021}. In the recent years, this approach was scaled to deep architectures~\cite{wetzstein2020} and has proven spectacularly powerful in the field of computer vision, by exploiting diffraction~\cite{lin2018} or by optical implementations of convolutional neural networks, both in free-space~\cite{miscuglio2020} and on chip~\cite{feldmann2021}. Such architectures are able to extract increasingly abstract representations~\cite{zeiler2014} of the input images fed into the network by subsequent applications of pooled optically-operated linear convolutions.
    
    The most standard neural-network-based machine learning schemes present roughly the following architecture: the digital data to be processed are transformed by a series of consecutive layers that consist in a parametrized affine transformation followed by the point-wise application of an elementary nonlinear activation function. Such a composition of parametrized functions is then expected to approximate some target function of the input upon a proper training process involving the optimization of the parameters in order to minimize the error of the model over some set of training examples. While this sequential architecture is ideally suited for standard processors, which offer arbitrary levels of programmability by design, the amount of parameters to be addressed during the training process is in practice a hurdle to flexible optical implementations, having progressed from roughly $63$ thousands in the paradigmatic LeNet-5 convolutional neural network~\cite{lecun1998} to several tens of millions on its nowadays counterparts~\cite{khan2020}.
    
    Therefore, new machine learning paradigms that relax the above deep-architecture constraints, such as extreme-learning machines~\cite{huang2006,huang2012}, echo-state networks or reservoir computing~\cite{jaeger2001,jaeger2004,maass2002,verstraeten2005,verstraeten2006,lukosevicius2009,rodan2011,tanaka2019,nakajima2020,marcucci2020}, have inspired theoretical proposals and experimental realizations in a variety of settings, in free-beam optics~\cite{sande2017,sunada2020,pierangeli2021}, integrated photonics~\cite{vandoorne2014,denis-lecoarer2018}, memristors~\cite{kulkarni2012,du2017} and beyond~\cite{boyn2017,nakane2018,markovic2019}. In the context of optics, many fruitful configurations have been investigated, such as delay-line-based setups~\cite{vandoorne2008,paquot2010,appeltant2011,vandoorne2011,duport2012,paquot2012,larger2012,soriano2013,nguimdo2015,ortin2015,larger2017,chembo2020}, nonlinear polariton lattices~\cite{opala2019,ballarini2020,mirek2021} and systems combining linear light scattering and the measurement nonlinearity~\cite{dong2020,rafayelyan2020,pierangeli2021}.
    Beyond liquid-state machines/echo-state networks, several proposals have been put forward to reconcile general-purpose machine learning and distributed memory architectures. This is the case of neuromorphic approaches, in particular in the field of spintronics~\cite{sengupta2016,grollier2016,torrejon2017,romera2018,grollier2020} and memristors~\cite{jo2010,yu2011,yang2015,kim2015,chu2015,prezioso2015,jeong2016,hsieh2016,wang2017,yoon2018,wang2018,li2018a,kim2018,jeong2018}, notably the works articulated around spike-based machine-learning schemes~\cite{afifi2009,querlioz2011,querlioz2013,srinivasan2016,roy2019,sengupta2019}. Such architectures come with their own technical difficulties. Because they rely on co-located memory and processing resources, standard optimization algorithms can prove impractical, although novel optimization procedures have recently been put forward to overcome this obstacle by meeting their peculiar hardware design \cite{ernoult2019,ernoult2020,laborieux2021,laydevant2021,martin2021}.
    
    Besides the limited processing throughput, the above physical machine-learning paradigms naturally address a second shortcoming of standard software machine learning that relates to applications involving analog data that cannot be suitably interfaced with digital processors. Instances of this problem are, for example, situations where the input data to be analyzed are supplied at a throughput too high to be properly sampled in real time. This also occurs when the data are intrinsically analog, or when direct measurement processes add noise or perturb the system being measured. The utmost example of this is provided by genuinely quantum tasks with no classical counterparts, that involve quantum inputs~\footnote{In the recent years, several works have also addressed this matter from a privacy perspective, proposing quantum-secure machine-learning protocols~\cite{bang2015,sheng2017,song2021,Li2021}}. This has stimulated many original works, ranging from quantum metrology~\cite{ghosh2019a,ghosh2020} and quantum state control~\cite{ghosh2019,ghosh2021,krisnanda2021} to classical image recognition tasks with quantum hardware~\cite{xu2021}, under the name of quantum neuromorphic computing \cite{markovic2020b}.
    
    Here, we take a different path and propose to implement a distinct machine-learning paradigm, namely kernel machines~\cite{boser1992,scholkopf2001}, in the world of photonics. The approach appears to circumvent the two difficulties listed above. We describe photonic kernel machines as well as the associated theoretical framework under very general assumptions. We explore the links between key concepts in support-vector-machine (SVM) theory and those of the machine proposed here. We show that, in contrast with general reservoir computing schemes, knowledge about the underlying internal representations may be revealed from measurable data, providing direct understanding of the learning process of actual hardware. By introducing a realistic physical model for a photonic kernel machine based on a two-dimensional lattice of coupled linear optical cavities, we numerically examine the performance of photonic kernel machines on regression and classification tasks involving ultrafast spectral analysis of analog and noisy radio-frequency (RF) signals, which are imprinted on an optical carrier wave.
	
	This paper is organized as follows. After presenting some general concepts of kernel-machine theory in Section~\ref{sec:2}, photonic kernel machines are theoretically introduced in Section~\ref{sec:3}; their learning mechanism is discussed therein. A model for a physical implementation based on a photonic lattice is then described in Section~\ref{sec:4}. This model is numerically simulated in Section~\ref{sec:5} and applied to the ultrafast spectral analysis of noisy radio-frequency signals from single-shot optical intensity measurements. Finally, conclusions are drawn in Sec.~\ref{sec:6}.
	
	\section{General framework\label{sec:2}}
	
	Supervised machine-learning problems can be formulated as an optimization problem in which one tries to best approximate a \emph{target} quantity $\vb*{y} = f(\vb*{x})$ of some given input $\vb*{x} = [x_1, x_2,\ldots]^T$ with a parametrized function $\hat{f}$. The input data are distributed according to some unknown distribution $p(\vb*{x})$ from which only a restricted set of examples $\lbrace(\vb*{x}^\pidx{i}, \vb*{y}^\pidx{i})\rbrace_{i=1}^N$ is known.
    
    More precisely, in \emph{regression} problems one seeks for the optimal parameters that make $\hat{f}$ the best fit for the known examples by approximating their unknown true functional dependence $f$. At variance, in classification problems one aims at determining a predictor $\hat{f}$ that best associates to any given input $\vb*{x}$ a set of labels $\vb*{y}$ that characterizes its belonging to one or more classes. These approximation problems are expressed in practice as the minimization of some cost function $J$ with respect to the parameters of the model to be trained, for a given set of examples.
    
    A specific choice of parametrization defines the architecture of the model. Many such architectures exist, ranging from shallow models~\cite{hastie2009}, such as support vector machines (SVM), tree-based models and reservoir computing, to deep-learning models~\cite{goodfellow2016}, such as feedforward, convolutional or recurrent neural networks. Here, we will exclusively consider shallow models, with a strong emphasis towards kernel machines, which fundamental concepts and results will be introduced below. These notions will be particularly useful in the understanding of the photonic kernel machine discussed in Sec.~\ref{sec:3} and the associated physical-implementation model presented in Sec.~\ref{sec:4}.
	
	\begin{figure*}[ht!]
    	\centering
    	\includegraphics[width=.9\textwidth]{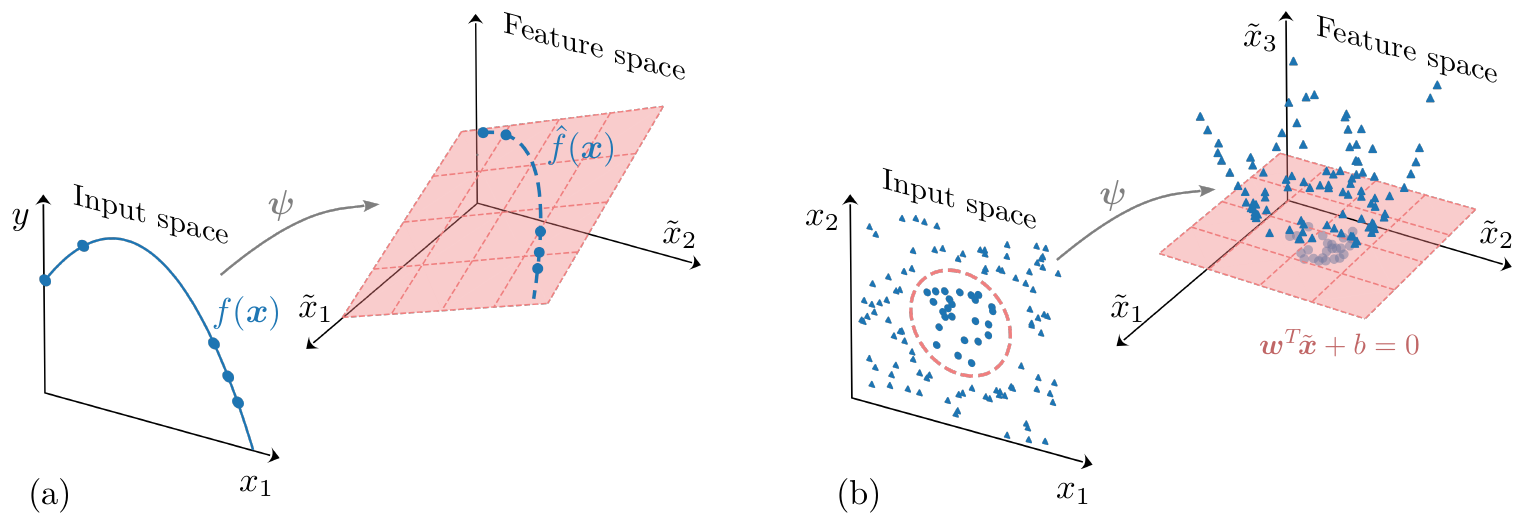}
    	\caption{Schematic representation of the embedding from input to feature space by a shallow model. (a) Example of regression: the quadratic function $f(x) = 3x - x^2 + 4$ is linearly fitted after the embedding of Eq.~\eqref{eq:1}, with $\psi_1(x) = x^2$ and $\psi_2(x) = x$. Note that, in feature space, all data points lie on a same plane. (b) Example of binary classification: triangles and dots become linearly separable in feature space after the embedding of Eq.~\eqref{eq:1}, with $\psi_1(\vb*{x}) = x_1$, $\psi_2(\vb*{x}) = x_2$ and $\psi_3(\vb*{x}) = x_1^2 + x_2^2$. The resulting input-space decision boundary $\hat{f}(\vb*{x}) = 0$ is represented by a dashed line. Origins are shifted to improve the legibility.}
        \label{fig:1}
    \end{figure*}

	\subsection{Shallow models\label{sec:2.A}}
	
	The principle underlying the learning process of shallow models is sketched in Fig.~\ref{fig:1}. Inputs are first embedded from their original \emph{input space} into a typically higher-dimensional \emph{feature space} where the optimization problem becomes linear. The optimization process is then realized therein, for instance by means of standard convex optimization algorithms, and often reduces to the geometrical problem of identifying an optimal hyperplane.
	
	More formally, the trial function $\hat{f}$ can be expressed in terms of a set of transformed (feature-space) coordinates  as
    \begin{equation}\label{eq:1}
    	\hat{f}(\vb*{x}) = \sum_{m=1}^{M}w_m \tilde{x}_m + b = \vb*{w}^T\tilde{\vb*{x}} + b,
    \end{equation}
    where
    \begin{equation}\label{eq:2}
    	\tilde{\vb*{x}} := \vb*{\psi}(\vb*{x}),
    \end{equation}
    with orthogonal components $\langle\psi_m,\psi_n\rangle_p = \mathbb{E}_p[\psi_m(\vb*{x})\psi_n(\vb*{x})] \propto \delta_{m,n}$, and $\mathbb{E}_p[\psi_m(\vb*{x})\psi_n(\vb*{x})] := \int\mathrm{d}\vb*{x}p(\vb*{x})\psi_m(\vb*{x})\psi_n(\vb*{x})$. Here, the \emph{feature map} $\vb*{\psi}: \vb*{x} \mapsto \tilde{\vb*{x}}$ defines an embedding from the \emph{input space} into a \emph{feature space} of dimension $M \leq +\infty$, with a new associated set of coordinates $\tilde{\vb*{x}} = [\tilde{x}_1, \ldots, \tilde{x}_M]^T$~\footnote{Note that this in general is not a diffeomorphism and thus \textit{stricto sensu} not a change of coordinates.}. Then, the trial function~\eqref{eq:1} bears the form of the equation of a hyperplane of parameters $(\vb*{w}, b)$ lying within this feature space.
    
    As shown in Fig.~\ref{fig:1}(a), a nonlinear function in input space can indeed become linear when expressed in a higher-dimensional feature space spanned by nonlinear transformations of the inputs. The feature-space embedding maps the original datapoints onto a hyperplane therein. The model of Eq.~\eqref{eq:1} can thus perform regression by approximating the target function as a plane parametric curve $\tilde{\vb*{x}} = \vb*{\psi}(\vb*{x})$ lying on this possibly infinite-dimensional hyperplane, whose parameters $(\vb*{w}, b)$ are to be determined.
	%
	
	
	Binary classification can be operated in a similar way. Inputs $\vb*{x}$ are associated to some class $a$ if $\hat{f}(\vb*{x}) > 0$ and to the complementary class $b$ otherwise. While this function might be involuted in input space, it may also become linear in some suitable feature space, the frontier between the two classes, $\hat{f}(\vb*{x}) = \vb*{w}^T\tilde{\vb*{x}} + b = 0$, thus becoming the equation of a plane. In this situation, the objective of the optimization becomes finding the proper plane in feature space that separates data belonging to the two distinct classes into two separate clusters, as illustrated in Fig.~\ref{fig:1}\,(b). This linear decision boundary translates back into a potentially non-trivial one in input space [dashed circle in Fig.~\ref{fig:1}\,(b)]. The same mechanism can be exploited to perform $k$-class classification via the one-vs-one and one-vs-rest techniques, which split the problem into respectively $k(k-1)/2$ and $k$ binary-classification problems.
    
    Such a shallow architecture presents a number of advantages. First, the optimization process is most often convex, in many cases analytic. Second, because it relies on a fixed feature-space embedding process, independent from any parameter to be updated during the training process, this architecture can be implemented on fast physical hardware, thus lifting any further analog-to-digital interfacing overhead.
	
	Many possible choices exist to realize the above feature-space embedding. One popular choice is that of reservoir computing~\cite{tanaka2019,nakajima2020}. This scheme exploits the nonlinear response of a dynamical system with many degrees of freedom---the \emph{reservoir}---to a driving input that encodes the data to be analyzed. As the set of produced independent nonlinear outputs for a single input is increased, the spanned feature space is expected to ``include'' the optimal one where the problem becomes linear~\cite{huang2006,huang2012}. While no control on the encoding of the inputs and the parameters of the reservoir is in principle required, in practice, this may impact the convergence of the performance of the model in the number of extracted nonlinear outputs. 
	
	In the following, a similar yet alternative approach will be considered, rooted in the theory of support-vector machines (SVM): the so-called \emph{kernel machines}. These rely on an educated guess : the feature-space embedding is induced based upon a choice of similarity metric, encoded in a \emph{kernel} $K$, which associates to any two given inputs $\vb*{x}$ and $\vb*{x}'$ a measure of their similarity $K(\vb*{x},\vb*{x}')$. Beyond providing heuristics on the way of building a feature-space embedding best suited for a specific task, such models come with theoretical bounds on the probability of generalization error~\cite{scholkopf2001,caponnetto2007,mohri2012}.
	
	In practice, kernel-method practitioners choose a kernel suitable for their applications; among popular choices are linear kernels, $K(\vb*{x}, \vb*{x}') = \vb*{x}^T\vb*{x}'$; polynomial kernels, $K(\vb*{x}, \vb*{x}') = (1 + \vb*{x}^T\vb*{x}'/c)^d$; radial basis functions (RBF)~\cite{powell1987,broomhead1988}, $K(\vb*{x}, \vb*{x}') = \exp(-\lVert\vb*{x} - \vb*{x}'\rVert_2^2/\sigma^2)$; or sigmoid kernels, $K(\vb*{x}, \vb*{x}') = {\tanh}(\kappa \vb*{x}^T\vb*{x}' + \theta)$; to name a few~\cite{vangestel2004}.
	Very recently, ``quantum'' kernels of the form $K(\vb*{x}, \vb*{x}') = \Tr[\hat{\rho}(\vb*{x})^\dagger \hat{\rho}(\vb*{x}')]$ or $K(\vb*{x}, \vb*{x}') = \langle\hat{A}(\vb*{x})^\dagger \hat{A}(\vb*{x}')\rangle$, where $\hat{\rho}(\vb*{x})$ and $\hat{A}(\vb*{x})$ denote quantum operators evolved through some input-dependent unitary transformation $\hat{U}[\vb*{x}]$ were proposed~\cite{liu2018,schuld2019,bartkiewicz2020} and even experimentally implemented~\cite{havlicek2019,bartkiewicz2020,kusumoto2021}. Quantum advantage on a classification task was demonstrated on such kernel machines~\cite{liu2020}; more recently, it was shown that circuit-based ``quantum neural networks'' could be described as such quantum kernel machines~\cite{schuld2021}.
	
	Although in principle large-scale data processing with kernels can be computationally very expensive, several approximate techniques have been developed to overcome this limitation~\cite{williams2001,drineas2005,rahimi2007,rahimi2009}, some of which have recently inspired physical implementations capable of efficiently performing approximate kernel evaluations of high-dimensional vectorial data (e.g. images) on optical hardware~\cite{saade2016,ohana2020}; see also Ref.~\cite{pierangeli2021} for interesting related strategies using free-space photonics.
	In the present work, we both define a kernel suitable for processing (infinite-dimensional) continuous time signals and provide a way to efficiently approximate it with photonic hardware.
	
	
	\subsection{Kernel machines}
	
	We here recall some fundamental identities from the theory of kernel machines that will prove useful in the following. 
	
	Upon making a specific choice of symmetric positive semi-definite kernel $K$, the kernel machine's predictions $\hat{f}$ can be evaluated in two different forms. First, as an average over a set of trainable model parameters $\vb*{\alpha} = [\alpha_1, \alpha_2, \ldots, \alpha_N]^T$, weighted over the similarity between the considered input $\vb*{x}$ and all those composing the training set, of size $N$:
	\begin{equation}\label{eq:3}
	    \hat{f}(\vb*{x}) = \sum_{i=1}^{N}\alpha_i K(\vb*{x},\vb*{x}^\pidx{i}) + b,
	\end{equation}
	where $b$ is a bias parameter. Alternatively, predictions can be obtained as an expansion over some set of $M\leq +\infty$ possibly-nonlinear functions $\lbrace h_m\rbrace_{m=1}^M$ such that $K(\vb*{x},\vb*{x}') = \sum_m h_m(\vb*{x})h_m(\vb*{x}')$, $\forall (\vb*{x}, \vb*{x}')$. Explicitly:
	\begin{equation}\label{eq:4}
        \hat{f}(\vb*{x}) = \sum_{m=1}^{M}\beta_m h_m(\vb*{x}) = \vb*{\beta}^T\vb*{h}(\vb*{x}),
    \end{equation}
    where $\vb*{\beta} = [\beta_1, \ldots, \beta_M]^T$, of size $M$, is the set of parameters of the model~\footnote{For simplicity, no bias is here considered, without loss of generality as one may always write $\hat{f}(\vb*{x}) = \sum_{m=1}^{M}\beta_m h_m(\vb*{x}) + b = \sum_{m=0}^{M}\beta_m h_m(\vb*{x})$, with $h_0 = 1$ and $\beta_0 = b$.}.
    
    The optimal parameters of the model ($\hat{\vb*{\alpha}}$ and $\hat{\vb*{\beta}}$, respectively) are determined by minimizing a cost function. Each approach leads to a different yet equivalent optimization problem, respectively known as the \emph{dual} and the \emph{primal} problem, of the form:
    \begin{equation}\label{eq:5}
        J(\vb*{\xi}) = \sum_{i=1}^N V\bigl(y^\pidx{i},\hat{f}(\vb*{x}^\pidx{i})\bigr) + \lambda R(\vb*{\xi}),
    \end{equation}
    with either $\vb*{\xi} = \vb*{\alpha}$ (dual problem) or $\vb*{\xi} = \vb*{\beta}$ (primal problem). Here, $V(\vb*{y}^\pidx{i},\hat{f}(\vb*{x}^\pidx{i}))$ is some pointwise error function on the predictions made by the model that one wants to minimize and $R$ a regularization term whose strength is controlled by means of a ``bias'' hyperparameter $\lambda$.
    
    Regularization prevents the trained model from overfitting the dataset, thus increasing its generalization power. In the following, ridge regularization will be considered, defined by
    \begin{equation}\label{eq:6}
        R(\vb*{\alpha}) = \frac{1}{2}\vb*{\alpha}^T\vb{K}\vb*{\alpha},\quad
        R(\vb*{\beta}) = \frac{1}{2}\lVert\vb*{\beta}\rVert^2;
    \end{equation}
    where $K_{ij} = K(\vb*{x}^\pidx{i},\vb*{x}^\pidx{j})$ denotes the \emph{kernel matrix}.
	
	Both the dual and the primal approaches can be shown to lead to a same feature-space embedding, of the form of Eq.~\eqref{eq:1}, with feature maps given by $\vb*{x} \mapsto \psi_m(\vb*{x}) = \sqrt{\gamma_m}\phi_m(\vb*{x})$, where $\gamma_m$ and $\phi_m$ simply derive from the kernel's eigendecomposition:
    \begin{equation}\label{eq:7}
        K(\vb*{x}, \vb*{x}') = \sum_{m=1}^M\gamma_m\phi_m(\vb*{x})\phi_m(\vb*{x}'),
    \end{equation}
	with $\gamma_{m+1} \leq \gamma_{m}$ and $\langle\phi_m,\phi_n\rangle_p = \mathbb{E}_{p}[\phi_m(\vb*{x})\phi_n(\vb*{x})] \equiv \int\mathrm{d}\vb*{x}p(\vb*{x})\psi_m(\vb*{x})\psi_n(\vb*{x}) = \delta_{m,n}$. This eigendecomposition can be given empirical estimates by approximating the actual probability density distribution of inputs $p(\vb*{x})$ by the \emph{empirical} one $\hat{p}(\vb*{x})$~\cite{williams2001}, such that $\int\mathrm{d}\vb*{x}\hat{p}(\vb*{x})f(\vb*{x}) = (1/N)\sum_i f(\vb*{x}^\pidx{i})$. As shown in the appendix, given a primal problem parametrized by $\lbrace h_m\rbrace_{m=1}^M$, the \emph{empirical eigenvalues} \smash{$\lbrace\hat{\gamma}_m\rbrace_{m=1}^{\mathrm{rank}(\vb{K})\leq M}$} are simply given by those of the matrix $\vb{k}/N = \vb{H}^T\vb{H}/N$, where $H_{im} = h_m(\vb*{x}^\pidx{i})$, with the following associated \emph{empirical eigenfunctions}:
	\begin{equation}\label{eq:8}
	    \hat{\phi}_m(\vb*{x}) = \frac{1}{\sqrt{N\hat{\gamma}_m}} \vb*{u}_m^T\vb*{h}(\vb*{x}),
	\end{equation}
	where $\vb*{u}_m$ denotes $\vb{k}$'s $m$th eigenvector. These altogether lead to the \emph{empirical feature map} $\hat{\psi}_m(\vb*{x}) = \sqrt{\hat{\gamma}_m}\hat{\phi}_m(\vb*{x})$.
	
\begin{figure}[ht!]
	    \centering
	    \includegraphics[width=.65\linewidth]{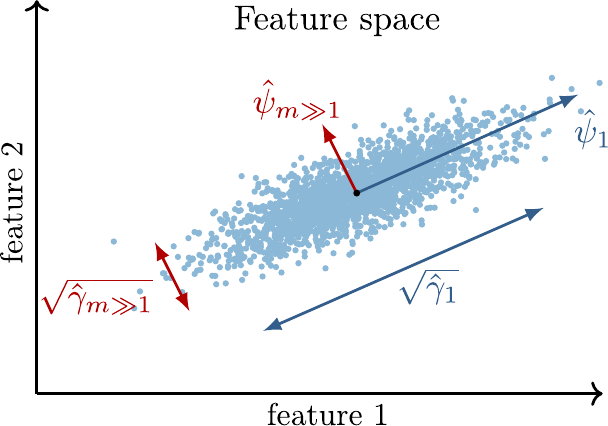}
	    \caption{Schematic representation of the projection of the feature-space-embedded data onto the plane spanned by the leading principal component of some kernel $K$ and some subleading one. The variance stemming from the correlations of the embedded training examples is expectected to be mostly contained within the dimensions spanned by the kernel's principal components; that of the noise, along orthogonal directions.}
	    \label{fig:2}
	\end{figure}

	The ability of kernel machines to filter out noise in input space can now be understood in very simple terms. To this aim, let us start from the primal approach and consider a kernel machine whose predictions are obtained as in Eq.~\eqref{eq:4}, with outputs centered by the empirical mean over the training set, \textit{i.e.} replacing $h_m(\vb*{x})$ by $\tilde{h}_m(\vb*{x}) = h_m(\vb*{x}) - \mathbb{E}_{\hat{p}}[{h}_m(\vb*{x})]$, where $\mathbb{E}_{\hat{p}}[X(\vb*{x})] = (1/N)\sum_i X(\vb*{x}^\pidx{i})$, for any $X$. It then follows that the corresponding kernel's empirical eigenfunctions are centered as well, $(1/N)\sum_i \hat{\phi}_m(\vb*{x}^\pidx{i}) = 0$, and the eigendecomposition~\eqref{eq:7} is in all respects analogous to principal component analysis on the feature-space-embedded data inputs~\cite{scholkopf1999,mika1999,scholkopf2001}. In this picture, schematically illustrated in Fig.~\ref{fig:2}, the empirical eigenvalues exactly correspond to the variance of the data along every feature-space direction:
	\begin{equation}\label{eq:9}
	    \hat{\gamma}_m = \mathbb{E}_{\hat{p}}\Bigl[\bigl(\hat{\psi}_m(\vb*{x}) - \mathbb{E}_{\hat{p}}[\hat{\psi}_m(\vb*{x})]\bigr)^2\Bigr].
	\end{equation}
	%
	%
	%
	%
	%
	Large variances of this type correspond to very correlated features that maximally deviate between inputs considered as ``dissimilar'' according to one's initial choice of similarity metric. Conversely, low variances correspond either to redundant features or to uncorrelated noise contained in the inputs.
	
	By now decomposing the trial function in the kernel basis identified above:
    \begin{equation}\label{eq:10}
    	\hat{f}(\vb*{x}) = \sum_m \langle\hat{\phi}_m,\hat{f}\rangle_{\hat{p}} \hat{\phi}_m(\vb*{x}) \equiv \sum_m \hat{f}_m \hat{\phi}_m(\vb*{x}),
    \end{equation}
    where the components $\hat{f}_m$ may be interpreted as degrees of freedom of the model, it follows from Eq.~\eqref{eq:6} that the regularization penalty can be rewritten as
    \begin{equation}\label{eq:11}
    	\lambda R(\vb*{\alpha}) = \sum_m\frac{\lambda_m}{2}\lvert\hat{f}_m\rvert^2,\quad
    	\lambda_m = \frac{\lambda}{\hat{\gamma}_m}.
    \end{equation}
    This means that the penalty on the magnitude of the $m$th component of $\hat{f}$ during the optimization is inversely proportional to the variance of the associated feature-space coordinate $\psi_m(\vb*{x})$.  
	Therefore, the effect of the regularization on the optimization is to impose a soft cut-off on the dimensionality of the feature-space hyperplane schematically illustrated in Fig.~\ref{fig:1}, while preserving most of the original variance contained in the inputs by only keeping the most statistically relevant directions, expected to encode the correlated information of interest in the inputs, while filtering out the noise~\cite{mika1999}. This can be understood as a filtering process as will clearly appear in the following.
	
	\section{Photonic kernel machines\label{sec:3}}

    The spectral analysis of radio-frequency signals is a very broad field with countless applications. We notably focus on the analysis of ultrashort pulsed signals. This is relevant for instance in the field of pulse-Doppler radars. Such devices emit a pulsed radio-frequency signal around a carrier frequency with a given repetition rate, and receive it back via an antenna after being reflected by a moving target. This reflected signal is measured and numerically processed to extract information about the position and the velocity of the tracked target. The repetition rate is limited by the processing time, which is in turn bounded by the acquisition time. Nowadays, the state of the art performances are obtained by direct sampling of the signal sensed by the antenna. Yet, the sampling rate of state-of-the-art RF analogue-to-digital converters lies typically below 10 gigasamples per second~\footnote{See for instance Texas Instruments' ADC12DL3200 ($6.4$\,GSPS) or Analog Devices' AD9213 ($10.25$\,GSPS)}. In order to measure $1000$ samples of a pulse of interest with such a converter, one needs the pulse length to exceed $\SI{25}{\us}$, limiting our ability to analyze shorter pulses. This issue may be circumvented by operating several such samplers in parallel. However, the resulting devices are very expensive, heavy and bulky.
    
    This context motivates the quest for technologies that go past these limitations. In what follows, we propose to resort to a learning all-optical processing device based upon the kernel-machine concept. We will first introduce a similarity kernel ideally suited for such spectral analysis tasks. We will show that this kernel can be evaluated rather naturally by means of an optical single-shot intensity measurement at the output of a photonic lattice. Such a photonic kernel machine makes it possible to process complex radio-frequency signals in times of the order of tens or hundreds of picoseconds.
    
    \begin{figure*}[ht!]
    	\centering
    	\includegraphics[width=.75\textwidth]{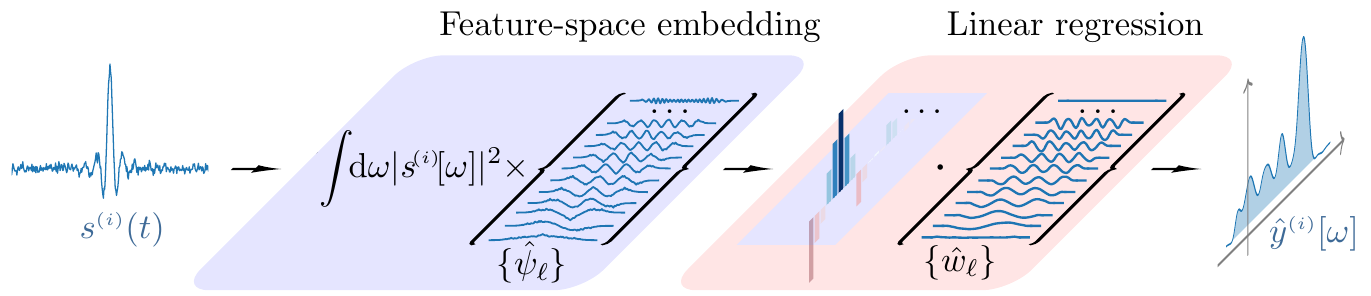}
    	\caption{Illustration of the photonic-kernel-machine processing mechanism. The measurement of the pulse-induced optical populations embeds the pulse's energy spectral density into some reciprocal space by projection onto a set of orthogonal base functions $\lbrace\hat{\psi}_\ell\rbrace_\ell$. The spectrum of the $i$th pulse is reconstructed by a linear combination $\hat{y}^\pidx{i}[\omega] = \sum_{\ell}w_{\ell}[\omega]\langle\hat{\psi}_{\ell}\vert S^\pidx{i}\rangle$ of its reciprocal-space components (features) over some set of learned functions $\lbrace \hat{w}_\ell\rbrace_\ell$. On this example, the learned functions $\lbrace \hat{w}_\ell\rbrace_\ell$ are filtered analogues of $\lbrace\hat{\psi}_\ell\rbrace_\ell$ and the photonic kernel machine is able to extract the spectrum of the incoming pulse from its noisy background.}
    	\label{fig:3}
    \end{figure*}
    
    \subsection{Theoretical description\label{sec:6.III.1}}
    
    As we have seen above, learning with kernels starts from the introduction of a kernel that serves as a measure of similarity between inputs. In particular, we here focus the discussion on the spectral analysis of ultrashort pulsed RF signals. The spectral information of such signals $s(t)$ is encoded in their energy spectral density $S[\omega] = \lvert s[\omega]\rvert^2$, where $s[\omega] = (2\pi)^{-1/2}\int\mathrm{d}t e^{-i\omega t}s(t)$. Rather naturally, the similarity between two signals $s(t)$ and $s'(t)$ can thus be given a general expression of the form:
    \begin{equation}\label{eq:12}
    	K(S,S') = \int\mathrm{d}\omega\mathrm{d}\omega' S[\omega]\mathcal{K}(\omega-\omega')S'[\omega'],
    \end{equation}
    where $\mathcal{K}(\omega - \omega')$ is a function peaked around $\omega = \omega'$ and with a typical width $\delta\omega$. Such a kernel compares two given input signals by contrasting their energy spectral densities at each frequency with a certain tolerance on their fine structure at frequency scales below $\delta\omega$. While this provides a good and flexible similarity metric, the numerical evaluation of such a kernel is rather unsuitable in practice using current electronics. Indeed, this would require several costly steps: (i) each pulse would have to be sampled in time at similar sampling rates over a time interval larger than $2\pi/\delta\omega$, (ii) the digitized signals would then have to be numerically Fourier-transformed and stored, (iii) finally, each kernel evaluation would involve calculating a double integral. In contrast, we will see that the feature-space embedding of this kernel can be performed rather naturally on photonic hardware. To that purpose, the RF signal to be analyzed can first be imprinted on an optical carrier injected into the hardware.
    
    Let us consider an optical system consisting of a set of $M$ generic normal modes as described by the set of bosonic operators $\lbrace\hat{\alpha}_\ell\rbrace_{\ell=1}^M$. Their response to any $i$th pulsed input signal $s^\pidx{i}(t)$ of interest is well described by a quantum Langevin equation~\cite{gardiner1985,gardiner2004}. In frequency space, this takes the form:
    \begin{equation}\label{eq:13}
    	\hat{\alpha}_\ell[\omega] = \chi_\ell[\omega] \Bigl(i s^\pidx{i}[\omega] + \sqrt{\kappa/2}\hat{\alpha}_\ell^\mathrm{in}[\omega]\Bigr),
    \end{equation}
    where $\hat{\alpha}_\ell[\omega] = (2\pi)^{-1/2}\int\mathrm{d}t e^{-i\omega t}\hat{\alpha}_\ell(t)$ is the Fourier-transformed annihilation operator associated to the $\ell$th normal mode, $\chi_\ell[\omega]$ its (model-dependent) susceptibility at frequency $\omega$, and $\kappa$ the loss rate, taken independent of $\ell$. Here, the Langevin input fields $\hat{\alpha}^\mathrm{in}_\ell(t)$ account for zero-temperature quantum noise and satisfy the usual expectation values: $\langle\hat{\alpha}_\ell^\mathrm{in}(t)\rangle = \langle\hat{\alpha}_\ell^{\mathrm{in} \dagger}(t)\hat{\alpha}_{\ell'}^\mathrm{in}(t')\rangle = 0$ and $\langle[\hat{\alpha}_\ell^\mathrm{in}(t),\hat{\alpha}_{\ell'}^{\mathrm{in} \dagger}(t')]\rangle = \delta_{\ell,\ell'}\delta(t-t')$.
    
    Optical populations in the modes, induced by the input pulse, are measured via the radiated optical power collected by a detector, with some integration time $\Delta t$ much larger than the length of the pulse, $\Delta t \gg 2\pi/\Delta\omega$, where $\Delta\omega$ is the bandwidth of the signal to be analyzed. In this limit, the measured normal-mode populations are expressed as
    \begin{align}
    	\bar{n}_\ell^\pidx{i} \simeq& \frac{1}{\Delta t}\int\mathrm{d}t \langle\hat{\alpha}_\ell^\dagger(t)\hat{\alpha}_\ell(t)\rangle = \frac{1}{\Delta t}\int\mathrm{d}\omega\langle\hat{\alpha}_\ell^\dagger[\omega]\hat{\alpha}_\ell[\omega]\rangle\nonumber\\
    	=& \int\mathrm{d}\omega h_\ell(\omega) S^\pidx{i}[\omega] \equiv \braket{h_\ell}{S^\pidx{i}},\label{eq:14}
    \end{align}
    where the integration window was approximately extended to $\mathbb{R}$ in the first equality, $h_\ell(\omega) = \Delta t^{-1}\lvert\chi_\ell[\omega]\rvert^2$ is the optical population susceptibility and $S^\pidx{i}[\omega] = \lvert s^\pidx{i}[\omega]\rvert^2$ the energy spectral density of the $i$th pulse.
    
    From such a population measurement, vector-valued predictions on any $i$th input energy spectral density $S$ take the form of those of a kernel machine as in Eq.~\eqref{eq:4}:
    \begin{equation}\label{eq:15}
    	\hat{f}(S) = \vb{B}^T\bar{\vb*{n}}(S),
    \end{equation}
    with components:
    \begin{equation}\label{eq:16}
        \hat{f}_n(S) = \sum_\ell B_{\ell n}\braket{h_\ell}{S},
    \end{equation}
    where $\vb{B}$ is a matrix of parameters, $S$ plays the role of the input, and the feature index $m$ in Eq.~\eqref{eq:4} is replaced by that of the optical modes $\ell$.
    %
    %
    For optical relaxation times ($2\pi\kappa^{-1}$) of the order of tens of picoseconds, such predictions can be realized for ultrashort RF pulses at a throughput above the tens of gigahertz.
    
    Let us check that the corresponding dual-picture kernel is indeed of the form of Eq.~\eqref{eq:12} under some general assumptions on the optical-mode density spectrum of the photonic system. Our considered kernel takes the form:
    \begin{align}
    	K(S,S') &= \bar{\vb*{n}}^T(S)\bar{\vb*{n}}(S') \equiv \sum_{\ell}\braket{S}{h_\ell}\braket{h_\ell}{S'}\nonumber\\
    	&= \int\mathrm{d}\omega\mathrm{d}\omega' S[\omega]\mathcal{K}(\omega,\omega')S'[\omega'],
    \end{align}
    with $\mathcal{K}(\omega,\omega') := \sum_\ell h_\ell(\omega)h_\ell(\omega')$ an integral kernel~\footnote{$K$ and $\mathcal{K}$ may be thought of as components of a same operator $\hat{K} = \sum_\ell \ketbra{h_\ell}$. Indeed, one has $K(S,S') = \bra{S}\hat{K}\ket{S'}$ and $\mathcal{K}(\omega,\omega') = \bra{\omega}\hat{K}\ket{\omega'}$}. Provided all normal modes share the same susceptibility, $h_\ell(\omega) = h(\omega-\omega_\ell)$, and for a continuous optical-mode density spectrum $\rho(\omega)$, this amounts to:
    \begin{equation}\label{eq:17}
    	\mathcal{K}(\omega,\omega') = \int \mathrm{d}\Omega\rho(\Omega)h(\omega-\Omega)h(\omega'-\Omega).
    \end{equation}
    By further assuming a smooth spectral density of width larger than the normal-mode linewidth, one has:
    \begin{equation}\label{eq:18}
    	\mathcal{K}(\omega,\omega') \simeq \rho\bigl(\tfrac{\omega + \omega'}{2}\bigr)\int \mathrm{d}t h(t)h(-t) e^{-i\lvert\omega-\omega'\rvert t}.
    \end{equation}
    The integral kernel is thus completely determined by the observable susceptibility, here the mode population via $h(\omega) = \lvert\chi[\omega]\rvert^2 / \Delta t$. For instance, the susceptibility of a linear optical mode $\chi_\ell[\omega] = 1/(-i(\omega-\omega_\ell) + \kappa/2)$ translates into a Lorentzian:
    \begin{equation}\label{eq:19}
    	\mathcal{K}(\omega, \omega') \propto \frac{1}{(\omega-\omega')^2 + \kappa^2},
    \end{equation}
    that was shown to perform better on some tasks than the more common \emph{radial basis function}~\cite{zhang2011}. This integral kernel indeed bears the desired form introduced in Eq.~\eqref{eq:12}, with a spectral tolerance $\delta\omega \sim \kappa$ given by the mode's linewidth.
    
    \subsection{Inspecting the feature space\label{sec:3.B}}
    
    One can access to the feature space associated to such a photonic kernel machine as well. Indeed, the above kernel admits an eigendecomposition analogous to Eq.~\eqref{eq:7}:
    \begin{equation}\label{eq:20}
    	K(S,S') = \sum_{\ell}\braket{S}{h_\ell}\braket{h_\ell}{S'} = \sum_{\ell}\gamma_\ell \braket{S}{\phi_\ell}\braket{\phi_\ell}{S'},
    \end{equation}
    with empirical eigenfunctionals and feature maps as given by Eq.~\eqref{eq:8}:
    \begin{equation}\label{eq:21}
    	\hat{\phi}_\ell(\omega) = \frac{1}{\sqrt{N\hat{\gamma}_\ell}}\vb*{u}_\ell^T \vb*{h}(\omega),\quad
    	\hat{\psi}_\ell(\omega) = \sqrt{\hat{\gamma}_\ell}\hat{\phi}_\ell(\omega),
    \end{equation}
    where $\vb*{u}_\ell$ and the empirical eigenvalues $\hat{\gamma}_\ell$ are obtained following the previous section as the $\ell$th eigenvector and eigenvalue of the matrix $\vb{k}/N = \vb{H}^T\vb{H}/N$, with now $H_{i\ell} = \braket{h_\ell}{S^\pidx{i}} = \bar{n}_\ell^\pidx{i}$. Strikingly, this kernel matrix, and thus the above feature maps, can be directly constructed in experiments from the optical-population measurements over the training set as
    \begin{equation}\label{eq:22}
    	\frac{1}{N} k_{\ell\ell'} = \frac{1}{N}\sum_i \bar{n}_\ell^\pidx{i} \bar{n}_{\ell'}^\pidx{i}.
    \end{equation}

    Now that the feature map is identified, let us examine how the geometric picture introduced in Sec.~\ref{sec:2.A} translates to the photonic kernel case. In particular, we consider a regression problem where the targeted quantity is a function of the angular frequency $y[\omega]$, as is the case, for instance, when trying to estimate the spectrum of a pulse ($y[\omega] = S[\omega]$). This scenario can be dealt with by frequency binning the target as $y_n = y[\omega_n]$ and trying to best fit it using an estimator of the form $\hat{y}^\pidx{i}[\omega_n] := \hat{f}_n(S^\pidx{i})$, for any pulse $i$. Then, it follows from Eqs.~\eqref{eq:16} and \eqref{eq:21}, that predictions can be rewritten in the form of Eq.~\eqref{eq:1}, as an expansion over feature-space coordinates:
    \begin{equation}\label{eq:23}
    	\hat{y}^\pidx{i}[\omega_n] = \sum_{\ell}\hat{w}_{\ell}[\omega_n]\langle\hat{\psi}_{\ell}\vert S^\pidx{i}\rangle
    \end{equation}
    where
    \begin{equation}\label{eq:24}
    	\hat{w}_{\ell}[\omega_n] = \sqrt{N}[\vb{u}^T\hat{\vb{B}}]_{\ell n},
    \end{equation}
    with $\vb{u} = [\vb*{u}_1, \ldots, \vb*{u}_M]$. Here, the quantities $\tilde{S}_\ell \equiv \langle\hat{\psi}_\ell\vert S\rangle$ play the role of feature-space coordinates [$\tilde{x}_m$ in Eq.~\eqref{eq:1}] and can be interpreted as reciprocal-space components of the input $S[\omega]$ with respect to the basis $\lbrace\hat{\psi}_\ell\rbrace_\ell$. The quantities $\hat{w}_\ell[\omega_n]$ correspond to the parameters of the feature-space hyperplane [$w_m$ in Eq.~\eqref{eq:1}] and can be interpreted as a set of ``functions'' $\lbrace\omega\mapsto \hat{w}_\ell[\omega]\rbrace_\ell$ learned during the optimization procedure. These may be evaluated explicitly at the chosen frequency bins $\lbrace\omega_n\rbrace_n$ from the measurement data and the trained parameters $\hat{\vb{B}}$ by making use of the above expression. This picture is schematically illustrated in Fig.~\ref{fig:3}, in analogy with Fig.~\ref{fig:1}(a), on the spectral-analysis task.
    
    Let us note that by setting the trainable parameters to $\hat{w}_{\ell}[\omega_n] = \hat{\psi}_{\ell}(\omega_n)/\hat{\gamma}_{n}$ in the absence of any optimization step, one directly has that $\hat{y}^\pidx{i}[\omega_n] = S^\pidx{i}[\omega_n]$. This means that a photonic kernel machine is, at least, able to reproduce the energy spectral density of a pulse $s(t)$ from a single-shot intensity measurement, provided the energy spectral density belongs to the linear span of $\lbrace\hat{\psi}_\ell\rbrace_\ell$, as will be indeed verified in Sec.~\ref{sec:5}.
    The action of the ridge regularization of Eq.~\eqref{eq:6} on such a spectral-analysis task can be analytically uncovered. Indeed, upon choosing mean square error as error function, $V(y, \hat{y}) = \sum_n \lvert y[\omega_n] - \hat{y}[\omega_n] \rvert^2$, the functions learned by the model can be shown to be given by
    \begin{equation}
        \hat{w}_{\ell}[\omega]\Bigr\rvert_{\lambda} = \frac{1}{1 + \lambda/\hat{\gamma}_{\ell}} \times \hat{w}_{\ell}[\omega]\Bigr\rvert_{\lambda=0},
    \end{equation}
    and are thus filtered by the photonic kernel machine in accordance with the inverse of the variance of their associated features over the training set:
    \begin{equation}
        \hat{\gamma}_\ell = \mathbb{E}_{\hat{p}}\Bigl[\bigl(\langle S\vert\hat{\psi}_\ell\rangle - \mathbb{E}_{\hat{p}}[\langle S\vert\hat{\psi}_\ell\rangle]\bigr)^2\Bigr].
    \end{equation}
    %
    %
    %
    
     \begin{figure*}[ht!]
    	\centering
    	\includegraphics[width=.82\textwidth]{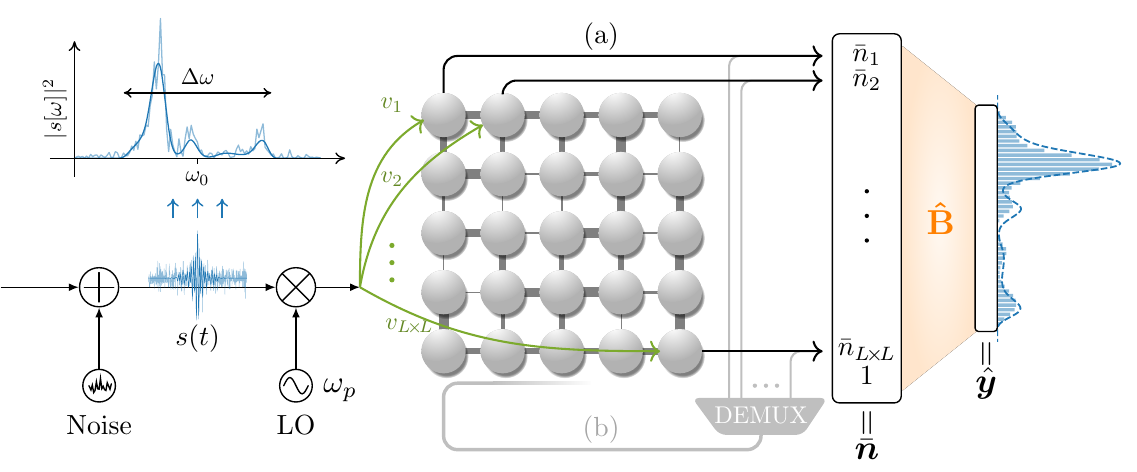}
    	\caption{\label{fig:4}%
    		Schematic representation of a trained photonic-lattice-based kernel machine estimating the energy spectral density of a noisy pulse from single-shot intensity measurements. The processing mechanism is the following: (i) A pulsed radio-frequency signal $s^\pidx{i}(t)$ is sent to the device in the presence of white noise; (ii) the noisy signal is then transferred to an optical carrier of angular frequency $\omega_p$ thanks to an electro-optic modulator ($\bigotimes$) and (iii) coherently injected into a lattice of $L \times L$ linear cavities with random parameters. (iv) The resulting optical populations are then measured and, finally, (v) the spectrum of the original noiseless signal is reconstructed by linearly combining the measured quantities thanks to the fixed matrix $\hat{\vb{B}}$, obtained from the previous training protocol. (a) Local-mode measurement protocol: populations are measured through the light locally radiated by the cavities. (b) Normal-mode measurement protocol: the cavities' optical field is first collected by an evanescently coupled waveguide and frequency-demultiplexed into $L\times L$ channels, whose optical populations are finally measured. Indeed, each normal mode has a different central frequency, allowing to spectrally separate the corresponding populations.
    	}
    	
    \end{figure*}
    
	\section{Physical implementation}\label{sec:4}
	
	We apply the above theory to the ultrafast processing of radio-frequency pulsed signals with a photonic lattice. The task consists in analyzing a set of baseband radio-frequency signals $\lbrace s^\pidx{i}(t)\rbrace$ over a bandwidth $\Delta\omega$ around some reference angular frequency $\omega_0$ by extracting some associated quantity of interest $\vb*{y}$, such as the energy spectrum or the peak frequency. To this aim, let us introduce an adapted physical implementation of a photonic-lattice-based kernel machine. The setup is schematically illustrated in Fig.~\ref{fig:4}. It involves four successive elements.

    Input signals enter the first unit of the system via an electro-optic modulator through frequency mixing with an optical carrier, $c(t) = c_0 \exp(-i\omega_p t)$, resulting in a modulated signal of the form $F^\pidx{i}(t) = s^\pidx{i}(t)e^{-i\omega_p t}$ whose central angular frequency $\omega_p + \omega_0$ may be shifted to accommodate the processing of signals in very different bands. The angular frequency of the local oscillator is set to $\omega_p = \bar{\omega} - \omega_0$, where $\bar{\omega}$ denotes the central angular frequency of the lattice's normal modes, to maximize the response of the system. The now optical modulated signal is then routed to the photonic lattice, entering the cavities as a coherent drive.
    
    The second unit consists of a $L\times L$ quadratic photonic lattice whose cavities are mutually coupled via near-field nearest-neighbor interactions. These cavities are coupled to the external modulated drive with some arbitrarily spatially-dependent weight $v_\ell$. In a frame rotating at the drive frequency, the dynamics of such a lattice is described by the following set of quantum Langevin equations~\cite{gardiner1985,gardiner2004}:
	\begin{align}
		\partial_t \hat{a}_\ell(t) &= \big[i(\Delta_\ell+\omega_0) - \kappa_\ell/2\big]\hat{a}_\ell(t) + i\sum_{\mathclap{\ell' \in N(\ell)}} J_{\langle \ell',\ell\rangle}\hat{a}_{\ell'}(t)\nonumber\\
		& + i v_\ell s(t) + \sqrt{\kappa_\ell/2}\;\hat{a}_\ell^\mathrm{in}(t),\label{eq:25}
	\end{align}
    where $\hat{a}_\ell$ is the photon annihilation operator for site $\ell$, $\Delta_\ell = \omega_p - \omega_\ell$ is the detuning of the local oscillator, $\kappa_\ell$ the optical relaxation rate, $J_{\langle \ell',\ell\rangle}$ the linear coupling rate between cavities $\ell$ and $\ell'$, $s(t)$ the broadband driving signal, and $v_\ell$ the local weight of the coupling of the $\ell$th cavity to the external drive. Operators $\hat{a}_\ell^\mathrm{in}(t)$ account for quantum noise and satisfy $\langle\hat{a}_\ell^\mathrm{in}(t)\rangle = \langle\hat{a}_\ell^{\mathrm{in} \dagger}(t)\hat{a}_{\ell'}^\mathrm{in}(t')\rangle = 0$ and $\langle[\hat{a}_\ell^\mathrm{in}(t),\hat{a}_{\ell'}^{\mathrm{in} \dagger}(t')]\rangle = \delta_{\ell,\ell'}\delta(t-t')$. A high degree of control of the parameters is not required. As a matter of fact, in the following numerical simulations, while angular frequencies $\omega_\ell$ will be uniformly set for simplicity, with $\Delta = -\omega_0$, the remaining parameters will be random variables. In particular, $J_{\langle m,\ell\rangle}$ will be uniformly drawn in the interval $[0, J_\mathrm{max}]$, $\kappa_\ell$ normally distributed around $\bar{\kappa} = zJ_\mathrm{max}/40$ ($z=4$) with a standard deviation of $10\%$, $\Delta\omega = zJ_\mathrm{max}$ and $\vb*{v}$ will be set to a normalized random real vector.
    
    The third unit consists of a sensor that measures the intensity of the light radiated by the optical cavities as they are externally driven by an input signal. These populations are measured by time-averaging over some detector integration time $\Delta t$ and collected into a vector $\bar{\vb*{n}}^\pidx{i} = [\bar{n}_1^\pidx{i}, \ldots, \bar{n}_{L{\times}L}^\pidx{i}, 1]^T$, where a unit entry is added to get a supplementary trainable parameter acting as a bias. In what follows, two alternative measurement settings will be considered:
    \begin{enumerate}[(i)]
    	\item Measure of the local populations: $\bar{n}_\ell^\pidx{i} = \frac{1}{\Delta t}\int\mathrm{d}t \langle\hat{a}_\ell^\dagger(t)\hat{a}_\ell(t)\rangle$. This corresponds, for instance, to the experimental situation where the intensity that is vertically emitted by the cavities is measured by a camera facing the lattice. This is schematically illustrated in Fig.~\ref{fig:4}\,(a).
    	
    	\item Measure of the normal-mode populations: $\bar{n}_\ell^\pidx{i} = \frac{1}{\Delta t}\int\mathrm{d}t \langle\hat{\alpha}_\ell^\dagger(t)\hat{\alpha}_\ell(t)\rangle$, where $\hat{\alpha}_\ell$ denotes the annihilation operator on the $\ell$th normal mode. This corresponds, for example, to the experimental situation where the field leaking from the cavities is collected by an evanescently coupled waveguide and frequency-demultiplexed into $L\times L$ frequency channels coupled to photodetectors. This is schematically illustrated in Fig.~\ref{fig:4}\,(b).
    \end{enumerate}
    
    Both of these settings are 
    
    Finally, a fourth unit performs a linear combination of the measured populations by acting with a $n \times (L^2+1)$ matrix of parameters $\vb{B}$, to be optimized by training. The output of this last unit is thus of the form $\vb*{\hat{y}}^\pidx{i} = \vb{B}^T\bar{\vb*{n}}^\pidx{i}$, as in Eq.~\eqref{eq:15}.
    
    \subsection{Training the lattice-based photonic kernel}
    
    The training starts from a training and a testing sets composed of $\Ntrain$ and $\Ntest$ signals, respectively, each of the form $\lbrace(s^\pidx{i}, \vb*{y}^\pidx{i})\rbrace_i$, where $s^\pidx{i}$ denotes the $i$th signal and $\vb*{y}^\pidx{i}$ a set of known associated features we want our system to learn how to estimate. In regression tasks, these labels $\vb*{y}^\pidx{i}$ may take arbitrary values, whereas in binary classification tasks $y^\pidx{i}=\pm1$ depending on whether the associated input $s^\pidx{i}$ belongs to a targeted class or not.
    
    The trial function of the untrained model is initially given by $\hat{f}(s^\pidx{i}) = \vb{B}^T\bar{\vb*{n}}^\pidx{i}$~\footnote{In practice, features $\bar{\vb*{n}}$ are first scaled using the z-score standardization prescription: $\bar{n}_\ell \mapsto (\bar{n}_\ell - \mathbb{E}_{\hat{p}}[\bar{n}_\ell]) / \sqrt{\mathrm{Var}_{\hat{p}}[\bar{n}_\ell]}$}. The training is carried out as follows:
    \begin{enumerate}[(i)]
    	\item \textbf{Construct the matrix $\vb{Y}$}. From the known features $\lbrace\vb*{y}^\pidx{i}\rbrace_i$ associated to the samples of the training set, $\vb{Y}$ is first computed as $Y_{im} = y_m^\pidx{i}$.
    	\item \textbf{Obtain the matrix $\vb{H}$}. Each input signal $s^\pidx{i}$ of the training set is fed into the device at the electro-optic modulator and the resulting cavity populations $\bar{\vb*{n}}^\pidx{i}$ are measured by the sensor. All these measured populations are stored in a matrix $H_{im} = \bar{n}_m^\pidx{i}$. Note that this evaluation is only to be performed once.
    	\item \textbf{Determine the optimal parameters $\hat{\vb{B}}$}. For a chosen value of the regularization hyperparameter $\lambda$, one obtains the optimal parameters by minimizing the cost function as \smash{$\hat{\vb{B}} = \argmin_{\vb{B}} J(\vb{B})$}, with \smash{$J(\vb{B}) = \sum_{i=1}^{\Ntrain} V(\vb*{y}^\pidx{i},\hat{f}(s^\pidx{i})) + \frac{\lambda}{2}\lVert\vb{B}\rVert_2^2$}, where $\hat{f}(s^\pidx{i}) = \vb{B}^T\bar{\vb*{n}}^{\pidx{i}}$, with $\bar{\vb*{n}}^{\pidx{i}}$ as obtained in the previous step. 
    	The choice of loss function $V$ depends on the task. For regression tasks, a popular choice is simply $V(\vb*{y}, \hat{\vb*{y}}) = \lVert \vb*{y} - \hat{\vb*{y}} \rVert^2$, which corresponds to a least-square problem; this directly yields $\hat{\vb{B}} = (\vb{H}^T\vb{H} + \lambda\vb{\mathds{1}})^{-1}\vb{H}^T\vb{Y}$ analytically, with $\vb{Y}$ and $\vb{H}$ as computed at steps (i) and (ii), respectively.
    	In binary classification ($y = \pm 1$), it is customary to choose a margin-maximizing loss functions~\cite{rosset2003}. Popular choices~\cite{hastie2009} are the hinge loss, $V(y, \hat{f}(s)) = \max(0,1 - y\hat{f}(s))$, which makes $\hat{f}$ directly approximate the class label $f(s) = y$, or the binomial deviance ${\ln}(1 + e^{-y\hat{f}(s)})$, which makes $\hat{f}$ instead approximate $f(s) = {\ln}(\mathbb{P}(y=+1\vert s)/\mathbb{P}(y=-1\vert s))$. Upon choosing any of these loss functions, the problem is convex and can be solved either analytically, for the square error, or by means of a convex optimization solver, using standard iterative methods.
    	
    	\item \textbf{Evaluate the accuracy of the model}. Once the model is trained, for any new input signal $s$ fed into the system, features are estimated according to $\hat{f}(s) = \hat{\vb{B}}^T\bar{\vb*{n}}$, from the measurement of the resulting populations $\bar{\vb*{n}}$. Its accuracy can then be benchmarked on the testing set by comparing the predictions $\vb*{\hat{y}}^\pidx{j} = \hat{f}(s^\pidx{j})$ against the known features $\vb*{y}^\pidx{j}$, for every input signal $s^\pidx{j}$ in the testing set, to which the model was yet never exposed. This is done through a metric that may differ from the cost function, in particular if regularization was employed.
    \end{enumerate}

    \subsection{Benchmarking the photonic-lattice kernel machine}
    
    In the following, two different approaches will be employed to benchmark the above-defined physical implementation of a photonic kernel.
    
    In order to evaluate the ability of the model to estimate the spectrum of pulsed signals, we will first use the modulus square of the fast Fourier transform (FFT) of the input signals as a reference of energy spectral density. We will assume an ideal sampling of the pulse over a centered window of time length $\Delta T = 5\times2\pi/\kappa$ at a sampling rate of $f_\mathrm{s} = 200 \times \kappa/2\pi$. For cavities with $2\pi/\kappa \sim \SI{10}{\ps}$, this corresponds to a sampling rate of $f_\mathrm{s} = \SI{10}{\THz}$. Note that this is more than three orders of magnitude beyond the state of the art~\cite{Note3}. Yet in the following the photonic kernel will be shown to outperform this rather fictional ideal device.
    
    We will also use a reservoir computing approach based on a nonlinear polaritonic lattice as first introduced in Ref.~\cite{opala2019}. This reservoir was numerically proven successful in image and speech recognition tasks and chaotic time-series forecasting, and was recently experimentally tested on an optical character recognition task~\cite{ballarini2020}. This reservoir is modelled by the following discrete complex Ginzburg-Landau equation:
    \begin{align}
    	\partial_t \alpha_\ell(t) &= \big[+ i(\Delta_\ell+\omega_0) + \gamma - (\Gamma + ig) \lvert\alpha_\ell(t)\rvert^2\big]\alpha_\ell(t)\nonumber\\
    	&+ i\sum_{\mathclap{\ell' \in N(\ell)}} J_{\langle \ell',\ell\rangle}\alpha_{\ell'}(t) + i v_\ell s(t),\label{eq:27}
    \end{align}
    where $\gamma = P - \kappa/2$ is a gain coefficient that accounts for the single-body decay rate $\kappa/2$ and the magnitude of the external pumping $P$, that brings the system close to instability. Here, $\Gamma$ and $g$ respectively account for two-body dissipation and interaction processes. The input signals are coherently injected in the same fashion as for the linear lattice. In the following numerical simulations, we will set $\Delta = -\omega_0$, $J_{\langle \ell',\ell\rangle}$ uniformly drawn in the interval $[0, J_\mathrm{max}]$, $\Gamma = zJ_\mathrm{max}/40$ ($z=4$), $\gamma/\Gamma = \num{8e-4}/2\pi$, $g/\Gamma = 1.6/2\pi$, $\Delta\omega = zJ_\mathrm{max}$ and $\vb*{v}$ to a normalized random real vector. The output of this model depends on the amplitude of the input. In what follows, the input energy will be set to $50\Gamma$.
    
    The training of this model is realized exactly as for the photonic kernel machine, from local population measurements $\bar{n}_\ell^\pidx{i} \propto\frac{1}{\Delta t}\int\mathrm{d}t \lvert\alpha_\ell (t)\rvert^2$.

    \begin{figure}[ht!]
    	\centering
    	\includegraphics[width=\linewidth]{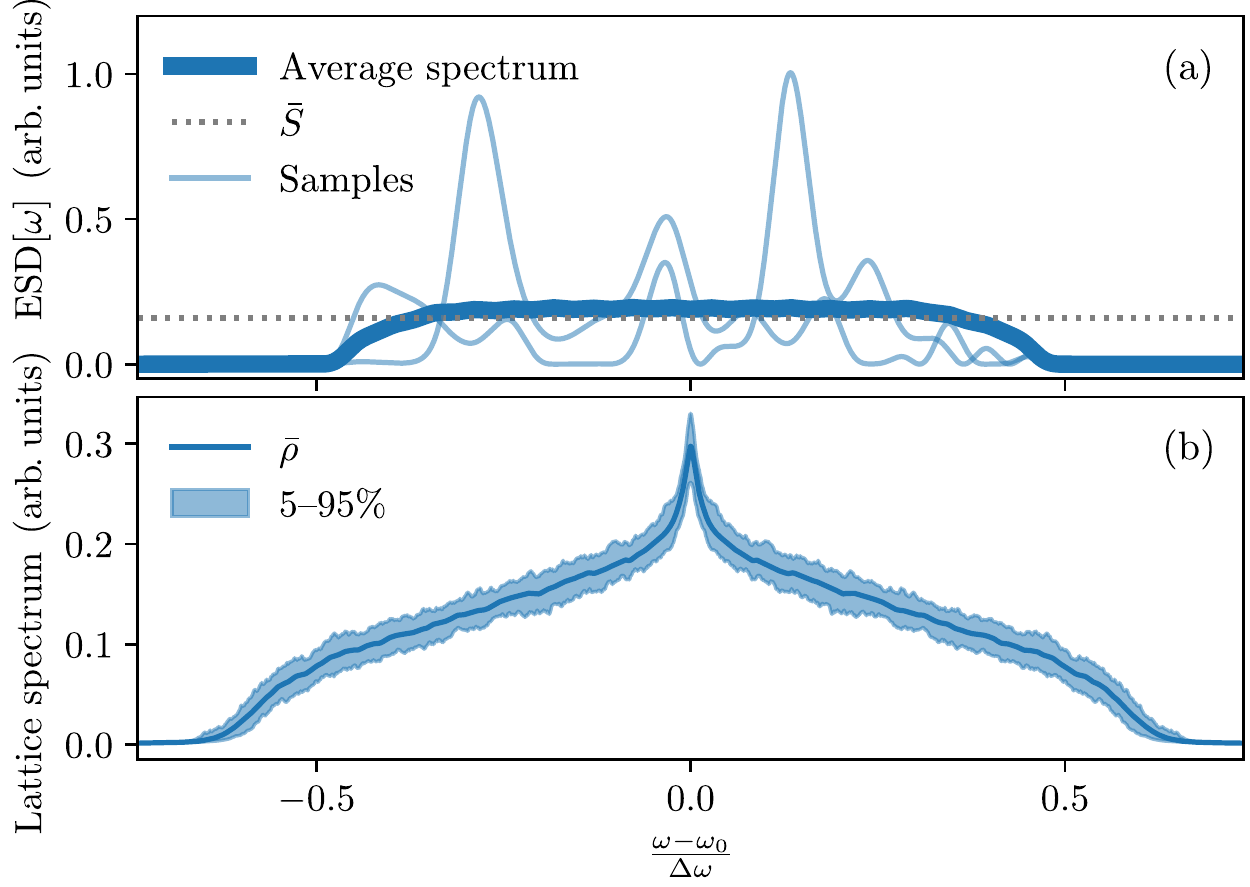}
    	\caption{%
    		(a) Energy spectral density averaged over $10^5$ realizations of the noiseless pulses, as given by Eq.~\eqref{eq:28}, as well as that of a few individual realizations. (b) Mean and interval between $5\%$ and $95\%$ quantiles for the optical spectra of $100$ realizations of a $30\times 30$ random photonic lattice.
    	}
    	\label{fig:5}
    \end{figure}
    \begin{figure}[ht!]
    	\centering
    	\includegraphics[width=\linewidth]{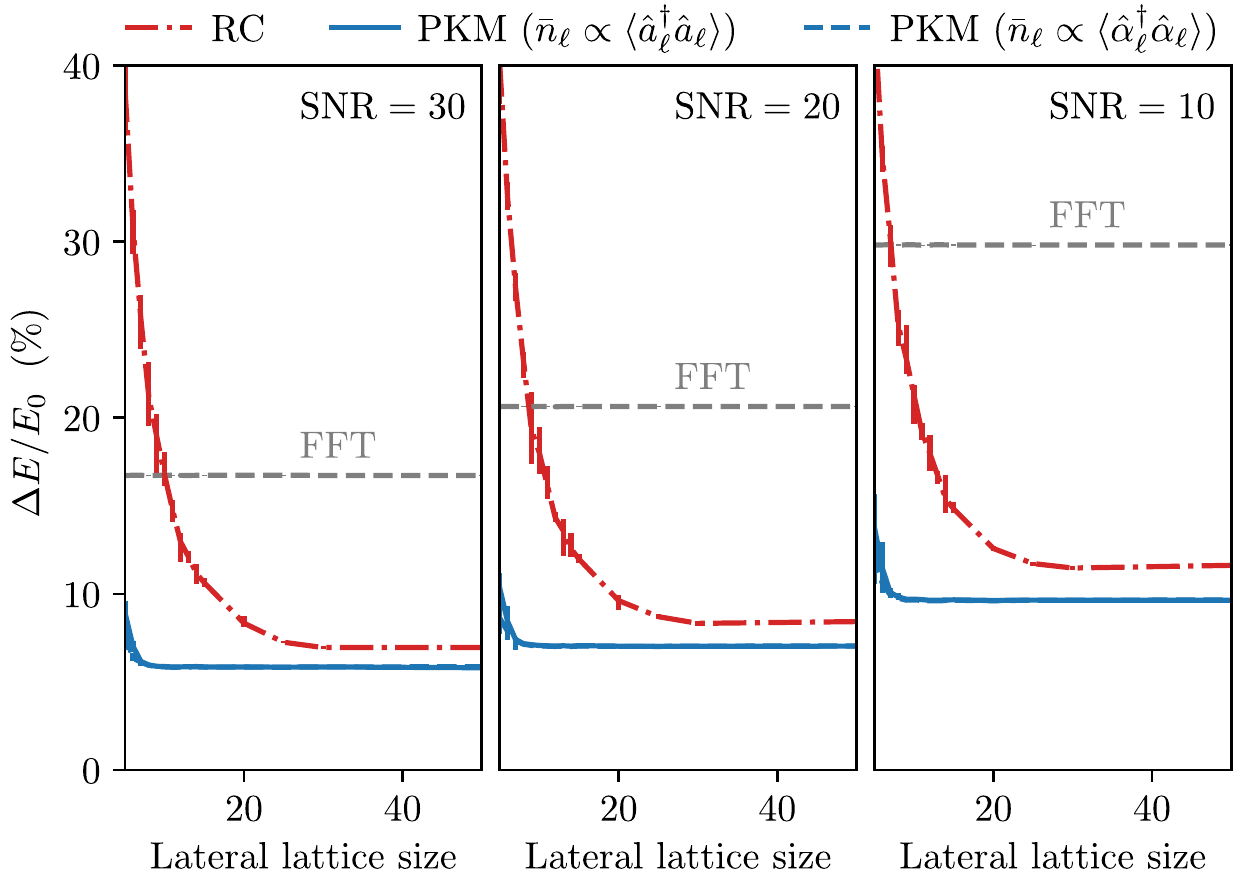}
    	\caption{%
    		Relative absolute error $\Delta E/E_0$ on the testing set for three values of the signal-to-noise ratio (SNR) for the fast Fourier transform (horizontal dashed line), the complex Ginzburg-Landau reservoir computing model (dash-dotted) and the photonic kernel machine where either the local-mode (plain) or the normal-mode (dashed) populations are measured. Data is averaged over $5$ realizations of the reservoir, error bars correspond to the intervals between the lowest and highest errors over these realizations. For each realization: $N_\mathrm{train} = 7000$ and $N_\mathrm{test} = 3000$.
    	}
    	\label{fig:6}
    \end{figure}
	
	\section{Applications\label{sec:5}}
	
	\subsection{Pulse spectral analysis}
	
    As a first illustration of the learning capabilities of the above-described photonic kernel machine, let us consider the extraction of the noiseless energy spectral density of an ultrashort pulsed radio-frequency signal embedded into a noisy background. Given an input noisy pulsed signal $s(t) + \xi(t)$, this task consists in giving the best possible estimation $\hat{y}_n$ of its frequency-binned energy spectral density $y_n = \lvert s[\omega_n]\rvert^2$ regardless of the noisy background $\xi(t)$.
    
    In order to train and study the performance of the model, we generate training and testing sets of $\Ntrain$ and $\Ntest$ pulses, respectively. To do so, we first generate known noiseless random spectra of same total energy $E_0 = \int\mathrm{d}\omega S_{ss}[\omega]$ by cubic B-spline interpolation of a set of $N_b$ signal bins $\tilde{s}_n := \tilde{s}[\omega_n]$, with frequency bins $\lbrace\omega_n\rbrace_n$ equally spaced within the band $[\omega_0 - \Delta\omega/2, \omega_0 + \Delta\omega/2]$ and randomly sampled from a Boltzmann distribution $p[\tilde{s}_n] = (1/Z) e^{-\beta V[\tilde{s}_n]}$ parametrized by the following potential:
    \begin{align}
    V[\tilde{s}_n] &= \frac{a}{N_b} \sum_{n=1}^{N_b-1} \lvert \tilde{s}_{n+1} - \tilde{s}_{n}\rvert^2 + b\big\lvert \max\limits_{n}\lvert\tilde{s}_n\rvert^2
    - S_\mathrm{peak}\big\rvert^2\nonumber\\
    &+ c \Big\lvert \std\limits_{n}\lvert\tilde{s}_n\rvert^2 - \bar{S}\Big\rvert^2,\label{eq:28}
    \end{align}
    with $\tilde{s}_i = 0$, $i=1,N_b$, $S_\mathrm{peak} = 8\bar{S}$, $\bar{S} = E_0/\Delta\omega$. Here, $a$ acts as stiffness parameter whereas the $b$-term favors peaked spectra. Finally, the $c$-term prevents the sampled spectra from sharing too similar shapes by favoring the presence of secondary peaks. The parameters used throughout this section are $N_b = 20$, $\beta a = \beta c = 100$ and $\beta b = 50$. The energy spectral densities of the obtained random noiseless spectra are shown in Fig.~\ref{fig:5}\,(a), where the average energy distribution is compared to $\bar{S}$ and a few typical examples are plotted, exhibiting various peaks with different heights. The choice of bandwidth ($\Delta\omega = z J_\mathrm{max}$) ensures that all the power of the signal to be analyzed can be sensed by the photonic lattice, as shown in Fig.~\ref{fig:5}\,(b), where the average optical spectrum of a large random photonic lattice is plotted. The random spectra are then Fourier-transformed to the time domain and white noise $\xi(t)$ is added to match some signal-to-noise ratio SNR, here defined as the relative contribution of the noise to the total energy in the analyzed band, i.e. \smash{$\mathrm{SNR} = E_0^{-1}\int_{\omega_0-\Delta\omega/2}^{\omega_0+\Delta\omega/2}\mathrm{d}\omega\lvert\xi[\omega]\rvert^2$}.
    
\begin{figure}[ht!]
    	\centering
    	\includegraphics[width=\linewidth]{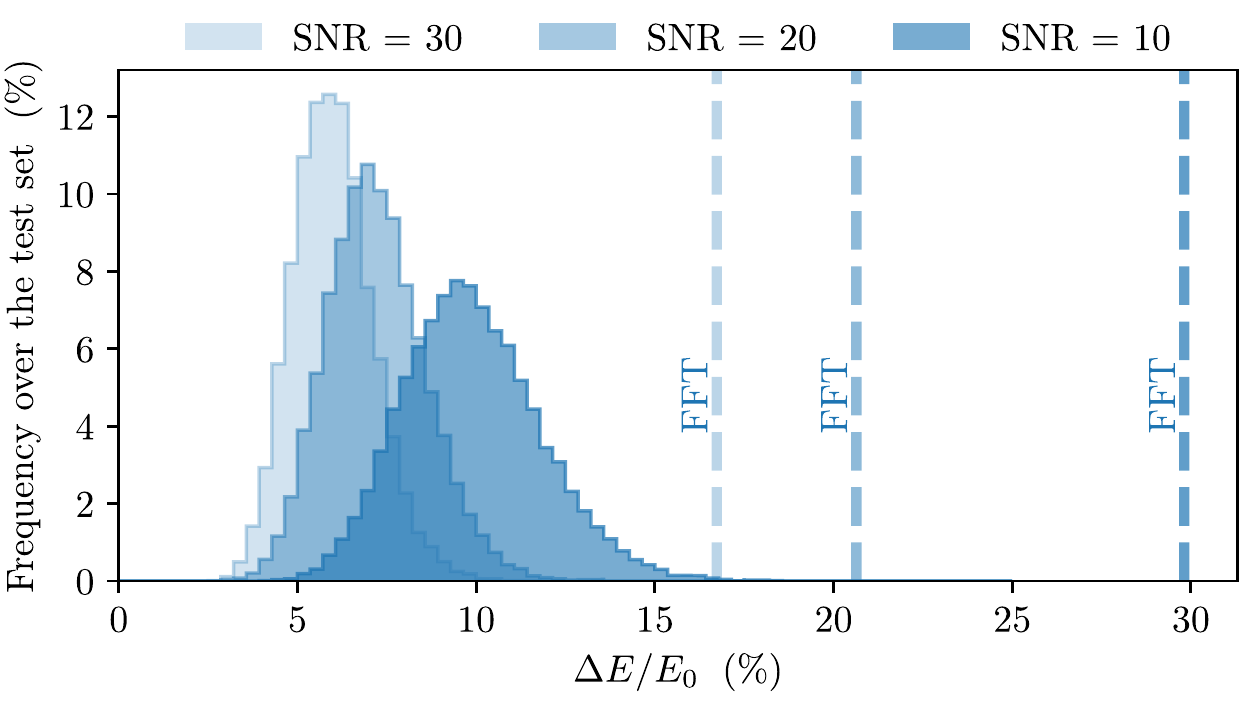}
    	\caption{%
    		Frequency of the error score over the testing set for three values of the signal-to-noise ratio averaged over $5$ realizations of a $20\times 20$ photonic kernel machine. $N_\mathrm{test} = 3000$ for each realization. The error score of the FFT is shown as dashed lines for comparison and falls systematically above the upper bound of that of the photonic kernel machine, in spite of being slower and requiring an ideal sampling rate significantly beyond the state of the art.
    	}
    	\label{fig:7}
    \end{figure}

    We then simulate the response of the photonic lattice to each of the driving input signals $s^\pidx{i}(t)$ of the training set by numerically integrating the coupled dynamical equations~\eqref{eq:25}. For each signal, either local or normal-mode populations are then measured yielding a set of time-averaged populations $\bar{\vb*{n}}^\pidx{i}$. The weights $\vb{B}$ are then optimized over the training data so as to minimize the square error between the predictions of the photonic kernel machine $\hat{\vb*{y}}$ and the known spectra of the noiseless pulses $\vb*{y}$. Upon fine-tuning of the hyperparameter $\lambda$ by 10-fold cross validation, the optimal weights $\hat{\vb{B}}$ are obtained analytically as explained above. The error of the trained model is then evaluated on the test set. We here quantify this error by the relative absolute error on the energy $\Delta E/E_0$, where $\Delta E$ represents the energy area between the estimated and the actual energy spectral density curves of the original noiseless spectra. In Fig.~\ref{fig:6}, we use this metric to benchmark our model against the ideal FFT of the noisy input signal and a the nonlinear polariton-based reservoir. For any of the chosen signal-to-noise ratio values and lattice sizes, the photonic kernel outperforms the other two approaches. Interestingly, the maximum performance is already reached with lattices as little as $10\times 10$, with an error in the reproduced spectrum about three times smaller than that obtained by the ideal FFT procedure. In addition, Fig.~\ref{fig:7} shows the error frequency over the testing set for the three values of the signal-to-noise ratio and a $20\times 20$ photonic lattice. Note that, for any amount of noise, the worst-case predictions remain more accurate than the ideal FFT. 
    
    \begin{figure}[ht!]
    	\centering
    	\includegraphics[width=\linewidth]{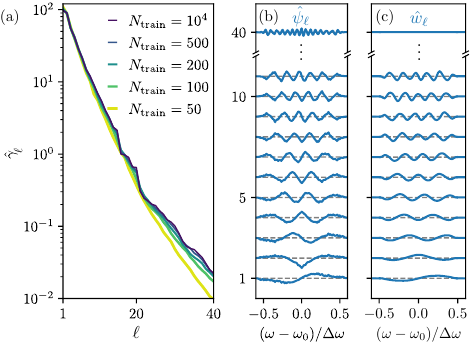}
    	\caption{(a) Convergence of the empirical spectrum of a photonic kernel for increasing training-set sizes. (b) First empirical eigenfunctions of the photonic kernel for $N_\mathrm{train} = 10000$. (c) Functions learned by the photonic kernel machine during the optimization process. Parameters are $L = 20$, $\mathrm{SNR} = 20$ and $\lambda = 10$.}
    	\label{fig:9}
    \end{figure}
    
    In order to understand how the device learns from training examples, let us make use of the theoretical concepts introduced above. As expressed in Eq.~\eqref{eq:23}, regression by the implemented photonic kernel machine is performed by a linear expansion $\hat{y}^\pidx{i}[\omega_n] = \sum_{\ell}\hat{w}_{\ell}[\omega_n]\langle\hat{\psi}_{\ell}\vert S^\pidx{i}\rangle$ of the feature-space components $\langle\hat{\psi}_{\ell}\vert S^\pidx{i}\rangle$ over a set of learned functions $\hat{w}_\ell$. The eigenvalues $\gamma_\ell$ and eigenfunctions $\psi_\ell/\sqrt{\gamma_\ell}$ of the photonic kernel do not depend on the optimization procedure and can be given empirical estimations from the measured populations $\bar{\vb*{n}}^\pidx{i}$ during the training process, as described above. In Fig.~\ref{fig:9}\,(a), we show the convergence of the empirical eigenvalues as the amount of training examples is increased, that is already reached for roughly $N_\mathrm{train} = 1000$ with our training protocol. Figs.~\ref{fig:9}\,(b) and (c) show the leading empirical feature maps as well as the corresponding learned eigenfunctions. On this task, one observes that the optimization procedure leads to a set of learned functions that correspond to filtered analogues of the empirical eigenfunctions of the kernel. One sees from Fig.~\ref{fig:9}\,(b) that the model builds a Fourier-sine expansion with a spectral resolution cut-off at $\sim\Delta\omega/2\pi L^2$. While in principle the optimization procedure may be sensitive to any feature-space component within this bound, the ridge regularization introduced during the optimization induces a soft cut-off for those whose associated eigenvalues have magnitudes lower than $\lambda$. This reduction to only the most statistically relevant components of the functional basis prevents the model from overfitting the training set, which would undermine its generalization capacity. In Fig.~\ref{fig:9}\,(c), the effect of the regularization is clearly visible on the highest-order represented learned function, that is completely filtered out. Hence, regularization here manifests itself as a low-pass filter on the learned decomposition.
    
    From this figure, the interpretation of the the photonic kernel machine regression mechanism becomes very clear: (i) the training set determines some optimal Fourier-like decomposition of the spectra; (ii) the regularization truncates the basis of such a decomposition to its most statistically relevant components, preventing the device from overfitting the noise; and, finally, (iii) the optimization finds a set of smooth filtered base functions on which to expand back those components to best reproduce the features. This results in the reconstruction of the noiseless spectrum, from which the noisy background uncorrelated with the training set is regularized out.

    \begin{figure}[ht!]
		\centering
		\includegraphics[width=\linewidth]{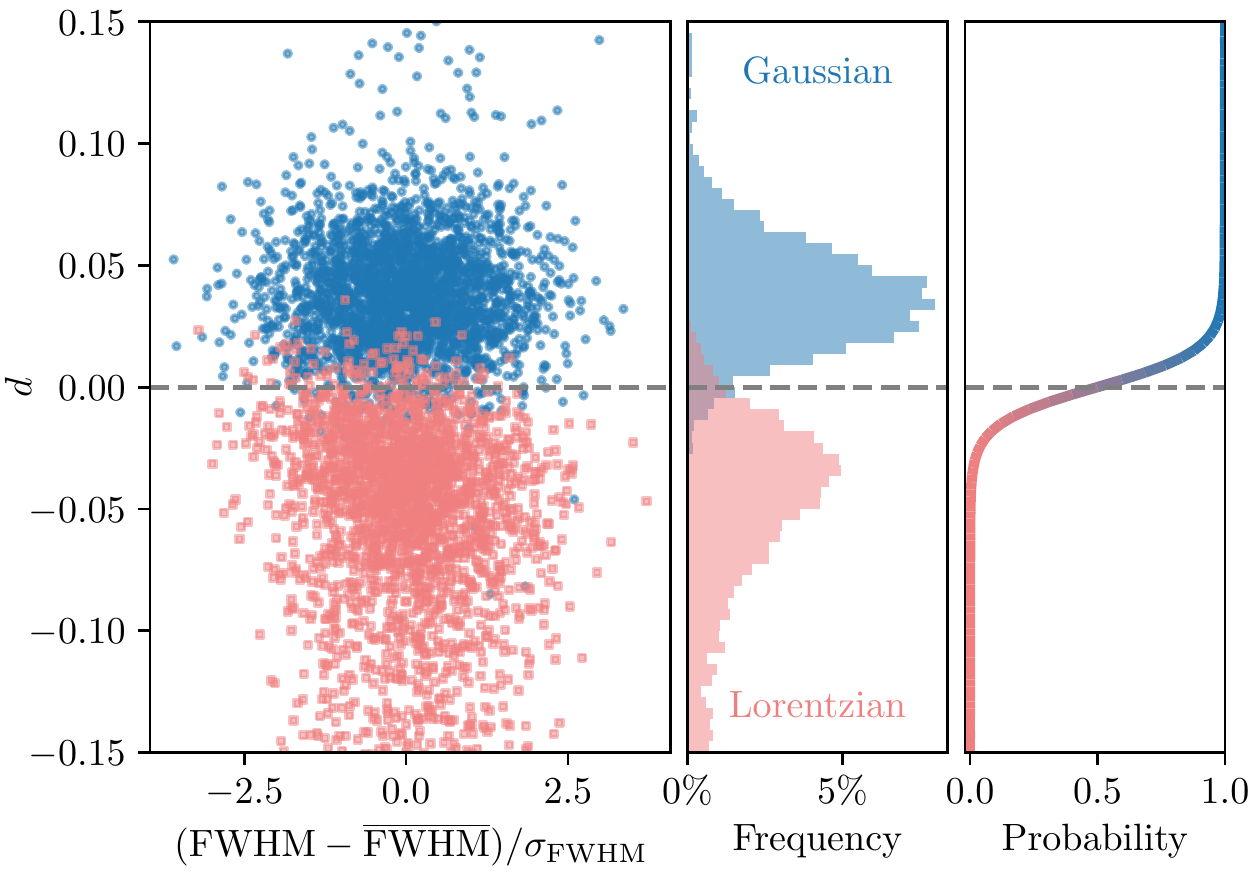}
		\vspace{0.2cm}
		\caption{(a) Signed distance from the testing pulses features to the learned hyperplane (decision boundary) as a function of their FWHM for a $20\times 20$ photonic kernel machine under normal-mode measurement of the photonic populations. Dots correspond to Gaussian pulses, squares to Lorentzian ones. Pulses falling above the decision boundary (dashed line) are categorized as Gaussian by the classifier. Parameters: $\mathrm{SNR} = 20$, $N_\mathrm{train} = 14000$ and $N_\mathrm{test} = 6000$. (b) Frequency of each class as a function of the signed distance to the discriminating hyperplane. (c) Probability that a pulse be Gaussian or Lorentzian at any given value of its associated signed distance to the discriminating hyperplane.}
		\label{fig:10}
	\end{figure}
	\begin{figure*}[ht!]
		\centering
		\begin{minipage}[t]{.69\textwidth}
			\strut\vspace*{-\baselineskip}\newline
			\includegraphics[width=0.8\linewidth]{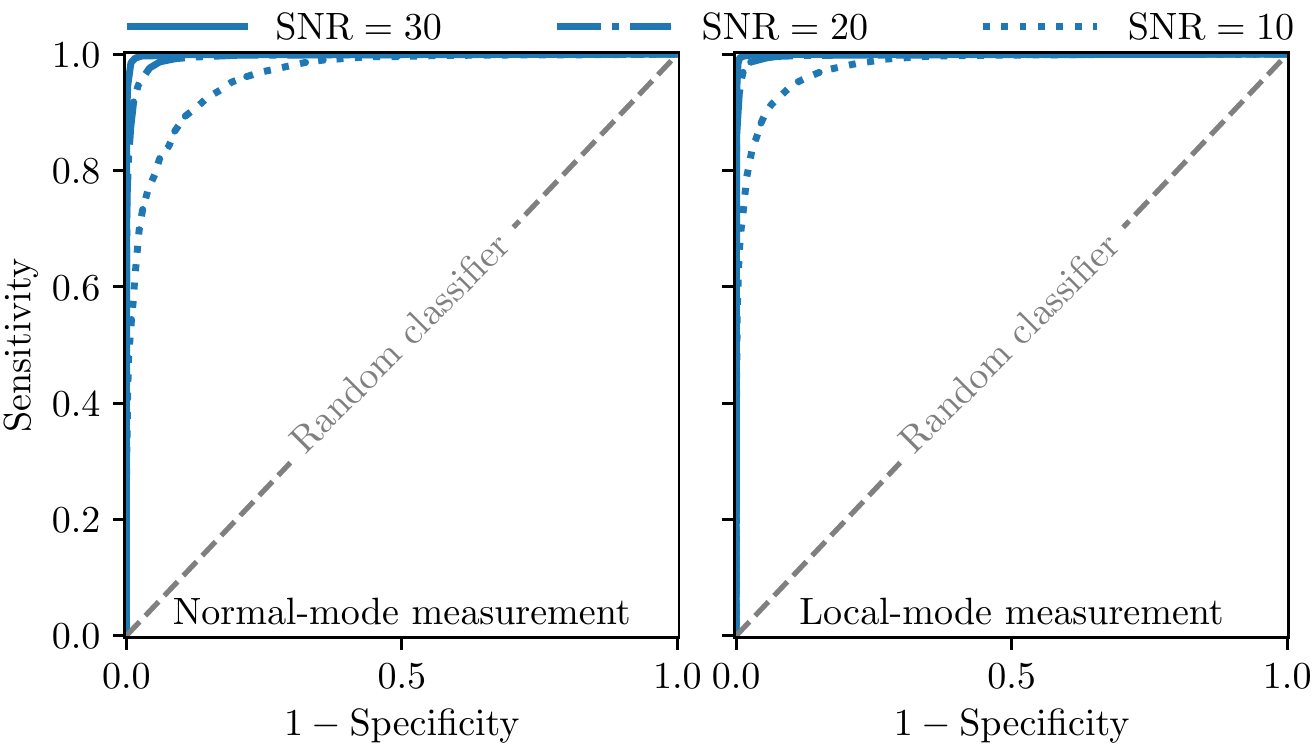}
		\end{minipage}
		\begin{minipage}[t]{0.3\textwidth}
			\strut\vspace*{-\baselineskip}\newline
			\includegraphics[width=0.9\linewidth]{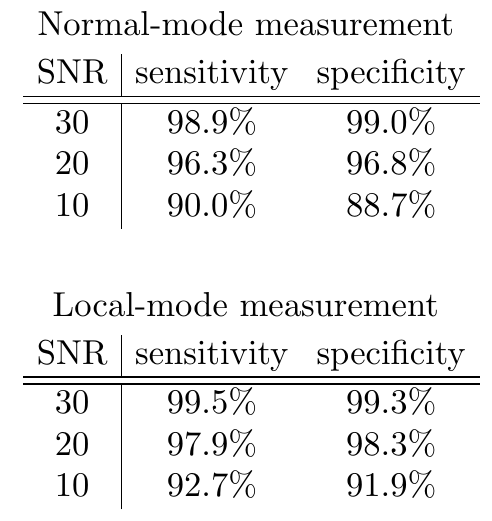}
		\end{minipage}
		\caption{Receiver operating characteristic (ROC) curve for a $20\times 20$ photonic kernel machine on the pulse shape classification task for three noise strengths and two measurement scenarios. The closer to the top left corner, the better. The numerical values of their associated sensitivities and specificities are compiled on the right tables. Parameters: $N_\mathrm{train} = 14000$ and $N_\mathrm{test} = 6000$.}
		\label{fig:11}
	\end{figure*}
	
	\subsection{Pulse shape recognition}
    
    We now show the performance of the same photonic kernel machine trained on a noisy-pulse shape classification problem.
    
    The task consists in determining whether the envelope of an input pulse is Gaussian ($y=1$) or Lorentzian ($y=-1$) from the measurement of the photonic populations of the lattice in the presence of noise at the input port of the system. We prepare a set of pulses with central angular frequencies uniformly drawn at random in the band of interest $[\omega_0-\Delta\omega/2,\omega_0+\Delta\omega/2]$ and full widths at half maximum (FWHM) normally distributed around $\overline{\mathrm{FWHM}} = \Delta\omega/20$ with a relative standard deviation of 10\%. The induced populations are then time-integrated over some detection time $\Delta t$ yielding a vector of intensities $\bar{\vb*{n}}$ that are finally linearly combined in such a way that the output of the photonic kernel machine is now a scalar of the form $\hat{f}(S) = \vb*{\beta}^T\bar{\vb*{n}}$. The optimization process is realized by minimizing the hinge loss \smash{$V(y^\pidx{i},\hat{f}(S^\pidx{i})) = \max(0,1 - y^\pidx{i}\hat{f}(S^\pidx{i}))$}. This is here achieved by means of a convex optimization solver~\cite{udell2014} for $\lambda \rightarrow 0^+$.
    
    The output of the $i$th pulse may be equivalently rewritten in terms of components of the feature map as $\hat{f}(S^\pidx{i}) = \hat{\vb*{w}}^T\tilde{\vb*{S}}^\pidx{i} + b$, with $\tilde{S}_\ell^\pidx{i} = \langle\psi_\ell\vert S^\pidx{i}\rangle$. It is worth noting that $\hat{f}(S) = \lVert\hat{\vb*{w}}\rVert \times  d(\tilde{\vb*{S}})$, that is, the prediction for any incoming pulse $s$ is proportional to the signed distance $d(\tilde{\vb*{S}})$ of its energy spectral density's feature-space coordinates to the plane ($\hat{\vb*{w}}$, $b$). As discussed above, this plane defines the feature-space decision boundary of the model. Input pulses are then classified into either of the two classes depending on whether their feature-space coordinates $\tilde{\vb*{S}}^\pidx{i}$ fall on either sides of this plane. Hence, predictions are of the form $\hat{y}^\pidx{i} = \mathrm{sign}(\hat{f}(S^\pidx{i}))$.
    
    The classification process is illustrated in Fig.~\ref{fig:10} for a $20\times 20$ photonic kernel machine and intermediate noise strength ($\mathrm{SNR}=20$). In panel~(a), one observes that features corresponding to either classes indeed cluster at either of the sides of the discriminating hyperplane independently from the value of the FWHM. As it becomes clear in panel~(b), predictions become less accurate as the distance from the separating hyperplane becomes smaller, thereby giving an estimation of the likelihood of the prediction. This can be made more quantitative by calibrating the probability of the classifier. This probability is shown in panel~(c) as given by $\mathbb{P}(s^\pidx{i}=\mathrm{``Gaussian"} \vert \hat{f}(S^\pidx{i})) = [1 + \exp(\smash[t]{-A\hat{f}(S^\pidx{i}) + B})]^{-1}$, where the calibration parameters $A$ and $B$ were determined by Platt scaling~\cite{platt1999}.
    
    The performance of a $20{\times}20$ trained photonic kernel machine is shown via its receiver operating characteristic (ROC) curve in Fig.~\ref{fig:11} for increasing noise strengths and the two population measurement scenarios. This displays the sensitivity (true positive rate) as one allows the specificity of the model to drop (higher false positive rates) by playing on some external bias added to the trained model, the best trade-off being found at the top left corner for no external bias. The ROC curve of an unbiased classifier that affects pulses randomly to either shape class is plotted as well for comparison. The high sensitivity and specificity values of the trained classifier are given in the right panel of Fig.~\ref{fig:11} for both population-measurement protocols.
    
    \subsection{Frequency tracking}
    
    Above, only the case of pulsed input signals was investigated. In the following, we illustrate the performance of the photonic kernel machine presented above on the analysis of continuous radio-frequency signals by considering frequency estimation of noisy sinusoidal radio-frequency signals. This is of great relevance for instance in the context of short-timescale force sensing with optomechanical devices~\cite{allain2020,guha2020}. In such applications, forces are read out by frequency tracking of the harmonic radio-frequency modulation imprinted on an optical signal by a resonating mechanical probe.
    
    To do so, we first generate a first set of training baseband signals consisting of complex exponentials with random initial phases and angular frequencies $\omega$ uniformly drawn between $\omega_0 - \Delta\omega/2$ and $\omega_0 + \Delta\omega/2$, to which white noise is added so as to match some signal-to-noise ratio $\mathrm{SNR}$, here defined as the ratio between the average power of the sinusoidal baseband signal and that of the noise. We then measure the steady-state populations of the cavity resulting from the driving of the coupled cavities by the modulated signals, here after a time $\tau_d = 10/\bar{\kappa}$, and use this vector of populations to make frequency predictions of the form $\hat{f}(s) = \vb*{\beta}^T\bar{\vb*{n}}$. The vector $\vb*{\beta}$ is then optimized so as to minimize the mean squared error between the estimated $\hat{\omega}^\pidx{i} = \hat{f}(s^\pidx{i})$ and the actual $\omega^\pidx{i}$ angular frequencies of all signals in the training set, with a ridge-regularization hyperparameter determined by 10-fold cross validation.
    
\begin{figure}[ht!]
    	\centering
    	\includegraphics[width=\linewidth]{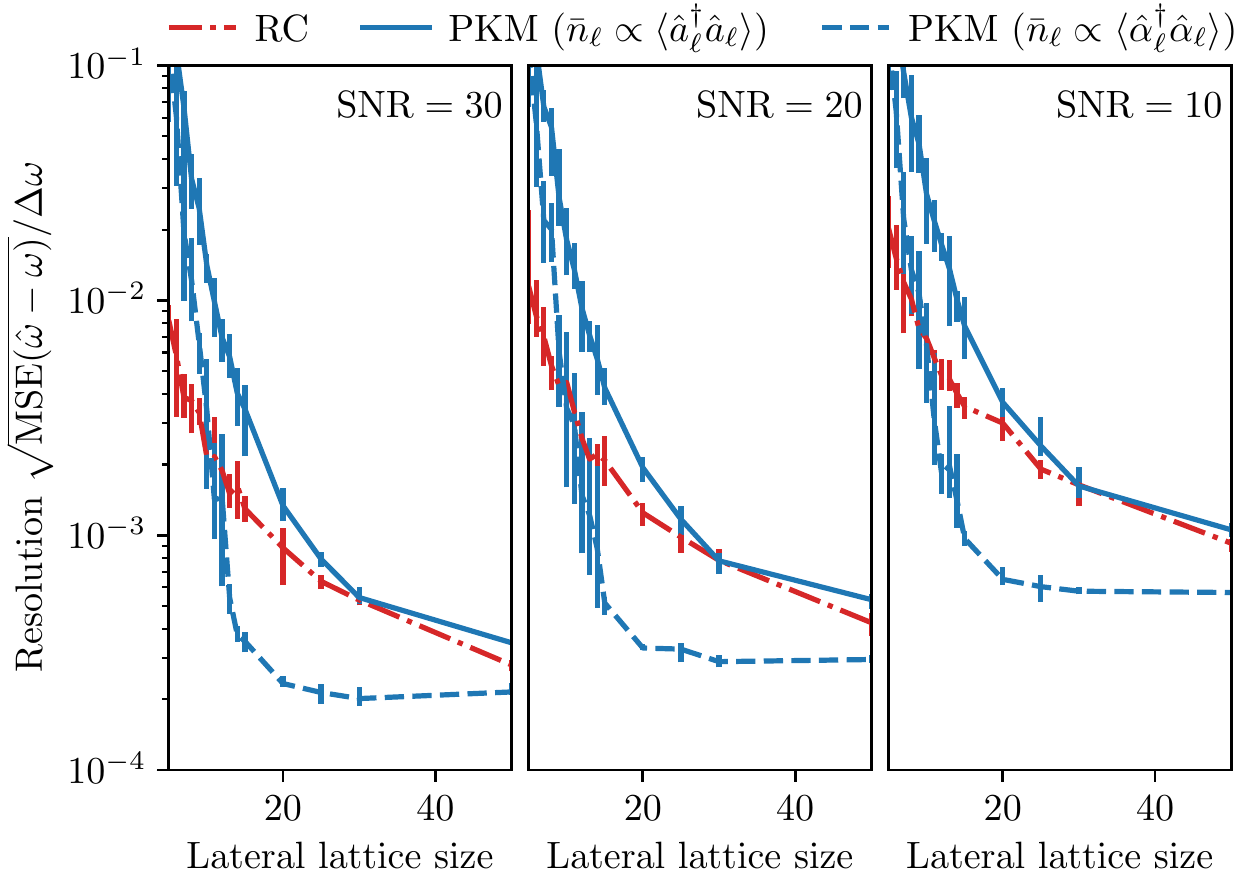}
    	\caption{%
    		Resolution of the photonic kernel machine on the testing set for either local-mode (plain) or normal-mode (dashed) population measurements for increasing lattice sizes $L$ and three values of the signal-to-noise ratio. The performance of the nonlinear polariton-based reservoir-computing scheme (dash-dotted) is shown for comparison. Data is averaged over $5$ realizations of the reservoir, error bars correspond to the intervals between the lowest and highest resolutions over these realizations. Parameters: $N_\mathrm{train} = 7000$ and $N_\mathrm{test} = 3000$.
    	}
    	\label{fig:12}
    \end{figure}

    The performance of the trained photonic kernel machine is finally evaluated on a testing set composed of new random complex exponentials. The achieved average resolution is shown in Fig.~\ref{fig:12} as a function of the lattice size for increasing values of the SNR, revealing the performance of the above-described photonic kernel machine on the spectral analysis of continuous radio-frequency signals. For cavities with $2\pi/\kappa \sim \SI{10}{\ps}$, the waiting time would be as little as $\tau_d \sim \SI{10}{\ps}$.
	
	\section{Conclusion\label{sec:6}}
    
    In this article, we presented photonic kernel machines, a framework for optical ultrafast spectral analysis of noisy radio-frequency signals that translates kernel methods from support-vector machines to photonic hardware. Such devices realize regression or classification tasks on high-dimensional data with throughputs above the gigahertz by utilizing the optical response of a set of optical modes to input analog signals as a measure of similarity.
    
    We first gave a theoretical description of photonic kernel machines under very general assumptions. We analytically investigated the similarity kernel built-in in such devices and were able to express it explicitly from the susceptibility of the measured observables. Furthermore, we explored the feature maps associated to photonic kernel machines and found that their expressions could be experimentally determined from population measurements and the knowledge of the single-mode susceptibility.
    
    We then studied a model describing a physical implementation consisting of a lattice of coupled linear optical cavities. We numerically demonstrated its capabilities on various regression and classification tasks, comprising the analysis of both pulsed and continuous radio-frequency signals. In particular, the proposed setup proved efficient in predicting the spectrum of picosecond pulses with nontrivial spectral structure from single-shot intensity measurements, being able to predict spectra with higher fidelity than the FFT of the noisy input signal. Moreover, it was shown to be able to discriminate pulses with distinct shapes as well as to estimate the angular frequency of continuous harmonic input signals. We showed that, by adding noise at the input of the device during the training protocol, the spurious effect of background noise on the predictive performance of the device could be successfully mitigated. On the spectrum estimation task, we could extract the actual feature maps associated to the simulated kernel machines as well as the basis of learned functions the photonic kernel machine composes its predictions from. This allowed us to interpret the photonic kernel machine regression mechanism and revealed the ability of the system to filter out the uncorrelated background noise.
    
    We believe that such devices, capable of analyzing above one million radio-frequency signals per second, may found applications in a broad variety of domains beyond spectroscopy. In the field of radio-frequency sensing, it could be used in pulse-Doppler radar systems as a way to optically analyze the reflected signals. In this way, the rate of emission could be increased by orders of magnitude, by relaxing the limiting dependence on the sampling rate of analog-to-digital converters. In telecommunications, these devices could be used, for instance, as a decoding means in FSK protocols involving high modulation rates in noisy environments. The fast frequency-tracking ability of the described device could be exploited in the context of short-timescale force sensing with optomechanical resonators. Finally, photonic kernel machines could be integrated into more conventional machine learning pipelines as a means of extracting non-trivial digital features from analog signals, acting as both an analog-to-digital converter and a preprocessing stage.
    

    
	
	
	
	\begin{acknowledgments}
	    This work was supported by the ERC Consolidator grant NOMLI No.~770933, by ANR via the project UNIQ, and the FET flagship project PhoQuS (grant agreement ID No.~820392).
	\end{acknowledgments}

	
	\appendix
	
	\section{Feature-space inspection}
    
    Kernel machines' feature space embedding may be straightforwardly examined. Indeed, for a symmetric positive semidefinite kernel $K$, Mercer's theorem ensures that it admits an eigendecomposition of the form:
    \begin{equation}\label{eq:a1}
    	K(\vb*{x}, \vb*{x}') = \sum_{m=1}^M\gamma_m\phi_m(\vb*{x})\phi_m(\vb*{x}'),
    \end{equation}
    with $\gamma_{m+1} \leq \gamma_{m}$ and $\langle\phi_m,\phi_n\rangle_{p(\vb*{x})} = \int\mathrm{d}\vb*{x}p(\vb*{x})\phi_m(\vb*{x})\phi_n(\vb*{x}) = \delta_{m,n}$, where the measure is set to the probability density function of inputs $p$. The feature map of Eqs.~\eqref{eq:1} and \eqref{eq:2} can then be easily shown to be given by $\psi_m(\vb*{x}) = \sqrt{\gamma_m}\phi_m(\vb*{x})$. We shall now identify the elements of this eigendecompositon.
    
    To identify its eigenvalues and eigenvectors, one has to solve in principle for
    \begin{equation}\label{eq:a2}
    	\int\mathrm{d}\vb*{x}'p(\vb*{x}')K(\vb*{x},\vb*{x}')\phi(\vb*{x}') = \gamma_m\phi_m(\vb*{x}).
    \end{equation}
    Yet these may be estimated from the $N$ known training samples by replacing the input distribution $p(\vb*{x})$ with the \emph{empirical} one $\hat{p}(\vb*{x})$, such that $\int\mathrm{d}\vb*{x}\hat{p}(\vb*{x})f(\vb*{x}) = (1/N)\sum_i f(\vb*{x}^\pidx{i})$~\cite{williams2001}. This yields a tractable discrete eigenvalue problem:
    \begin{equation}\label{eq:a3}
    	\frac{1}{N}\sum_{i=1}^N K(\vb*{x}, \vb*{x}^\pidx{i})\hat{\phi}_m(\vb*{x}^\pidx{i}) = \hat{\gamma}_m\hat{\phi}_m(\vb*{x}),
    \end{equation}
    in terms of \emph{empirical eigenvalues} $\hat{\gamma}_m$ and \emph{empirical eigenfunctions} $\hat{\phi}_m$. It follows that the empirical eigenvalues correspond to the non-zero eigenvalues of either of the kernel matrices $\vb{K} = \vb{H}\vb{H}^T$ and $\vb{k} = \vb{H}^T\vb{H}$ (let us recall $H_{im} = h_m(\vb*{x}^\pidx{i})$). Indeed, because of their Gram matrix structure, their eigendecompositions,
    \begin{equation}\label{eq:a4}
    	\vb{K} = N\vb{U}\vb{D}_{\hat{\gamma}}\vb{U}^T
    	,\quad
    	\vb{k} = N\vb{u}\vb{d}_{\hat{\gamma}}\vb{u}^T,
    \end{equation}
    with $U_{im} = \hat{\phi}_m(\vb*{x}^\pidx{i})/\sqrt{N}$, share the same non-zero eigenvalues and can be related through the following simple algebraic identity:
    \begin{equation}\label{eq:a5}
    	\vb*{U}_m = \tfrac{1}{\sqrt{N\hat{\gamma}_m}}\vb{H}\vb*{u}_m,
    \end{equation}
    following $\vb{K}$'s singular-value decomposition. In the physical model of the main text, one has typically $M\ll N$ and it becomes more suitable to work with the $M\times M$ matrix $\vb{k}$.
    
    Similarly, the empirical eigenfunctions can be determined from Eq.~\eqref{eq:a3}. By making use of the property of Eq.~\eqref{eq:a5}, these finally read:
    \begin{equation}\label{eq:a6}
    	\hat{\phi}_m(\vb*{x}) = \frac{1}{\sqrt{N\hat{\gamma}_m}}\vb*{u}_m^T\vb*{h}(\vb*{x}).
    \end{equation}
    
    From Eq.~\eqref{eq:a1} and Eq.~\eqref{eq:a6}, the kernel can be expanded into a series of empirical eigen feature maps $K(\vb*{x},\vb*{x}') = \sum_{m=1}^M \hat{\psi}_m(\vb*{x})\hat{\psi}_m(\vb*{x}')$, with orthogonal empirical feature maps simply given by
    \begin{equation}\label{eq:a7}
    	\hat{\psi}_m(\vb*{x}) = \sqrt{\hat{\gamma}_m}\hat{\phi}_m(\vb*{x}).
    \end{equation}
    Thus, the internal functional representations of the model are \emph{independent} from the optimization process, completely determined by the statistics of the training samples and can be estimated by diagonalizing the matrix $\vb{k}$.
    
    The predictions of the model in the abstract feature-space picture, $\hat{f}(\vb*{x}) = \vb*{w}^T\tilde{\vb*{x}} \equiv \vb*{w}^T\vb*{\psi}(\vb*{x})$, take the form of a linear combination of features whose weights, determined through training, can be geometrically interpreted as the parameters of a hyperplane. As a function of the primal parameters, this learned hyperplane is characterized by
    \begin{equation}\label{eq:a8}
    \hat{\vb*{w}} = \sqrt{N}\vb{u}^T\hat{\vb*{\beta}}.
    \end{equation}
    Therefore, provided a set of generating functions $\lbrace h_m\rbrace_m$, one is now able to access to a complete understanding of the model and its feature-space representations.

	\bibliography{references}
	
\end{document}